%% file: main.tex
\documentclass[lettersize,journal]{IEEEtran}
\usepackage{amsmath,amsfonts}
\usepackage{algorithm}
\usepackage{algpseudocode}
\usepackage{array}
\usepackage{textcomp}
\usepackage{stfloats}
\usepackage{url}
\usepackage{verbatim}
\usepackage{graphicx}
\usepackage{cite}
\usepackage{booktabs}
\usepackage{multirow}
\usepackage{makecell}
\usepackage{adjustbox}
\usepackage{xcolor}
\usepackage{arydshln}
\usepackage{amsthm}
\usepackage{thmtools} 
\usepackage[caption=false]{subfig}

\definecolor{estbg}{RGB}{229,236,246}
 
\makeatletter
\newcommand{\HState}[1]{\State{\setlength{\fboxsep}{1.5pt}%
  \hspace*{-\fboxsep}\colorbox{estbg}{\parbox{\dimexpr\linewidth-\ALG@thistlm-\fboxsep\relax}{#1}}}}
\makeatother

\theoremstyle{plain}
\newtheorem{theorem}{Theorem}[section]
\newtheorem{lemma}[theorem]{Lemma}
\newtheorem{corollary}[theorem]{Corollary}

\DeclareMathOperator{\Tr}{Tr}
\DeclareMathOperator{\diag}{diag}
\DeclareMathOperator*{\argmax}{arg\,max}
\DeclareMathOperator*{\argmin}{arg\,min}
\newcommand{\E}{\mathbb{E}}
\newcommand{\R}{\mathbb{R}}
\newcommand{\N}{\mathcal{N}}
\newcommand{\W}{\mathcal{W}}
\newcommand{\I}{\mathcal{I}}
\newcommand{\T}{\mathsf{T}}

\begin{document}

\title{AdaptivePnP: Fast Auto-Tuned Plug-and-Play for Deconvolution}

\author{Haley Duba-Sullivan ~\IEEEmembership{Student Member,~IEEE,} Emma J. Reid ~\IEEEmembership{Member,~IEEE,} Charles A. Bouman ~\IEEEmembership{Fellow,~IEEE,} and Gregery T. Buzzard ~\IEEEmembership{Senior Member,~IEEE}
\thanks{Haley Duba-Sullivan is with the Department of Mathematics, Purdue University, West Lafayette, IN 47907, USA and the National Security Sciences Directorate, Oak Ridge National Laboratory, Oak Ridge, TN 37830, USA (email: sullivanhe@ornl.gov).}
\thanks{Emma J. Reid is with the National Security Sciences Directorate, Oak Ridge National Laboratory, Oak Ridge, TN 37830, USA (email: reidej@ornl.gov).}
\thanks{Charles A. Bouman is with the School of Electrical and Computer Engineering, Purdue University, West Lafayette, IN 47907, USA (email: bouman@purdue.edu).}
\thanks{Gregery T. Buzzard is with the Department of Mathematics, Purdue University, West Lafayette, IN 47907, USA (email: buzzard@purdue.edu).}
\thanks{This work has been submitted to the IEEE for possible publication. Copyright may be transferred without notice, after which this version may no longer be accessible.}}



\maketitle

\begin{abstract}
Plug-and-play (PnP) methods reconstruct images by alternating a data-fitting step with an image denoiser. 
Their performance depends on two parameters: the step size $\gamma$ and the measurement noise level $\sigma$.
In practice, $\gamma$ is often tuned by grid search over validation images and must be retuned whenever the problem setting changes, and $\sigma$ is assumed known or estimated separately. 
We propose AdaptivePnP, a PnP framework that jointly estimates $\sigma$ and $\gamma$ at every iteration from the data-fitting residual.
To do so, we apply the PnP virtual model to the data-fitting proximal map and show that the residual is zero-mean Gaussian with covariance parameterized by $\gamma$ and $\sigma$. 
However, when one parameter is much larger than the other, the residual carries little information about the smaller one and using the maximum likelihood (ML) estimator can degrade the PnP iterations.
We therefore introduce the identifiable maximum likelihood (IML) estimator, which restricts the ML estimator to an identifiable window in which the normalized Cram\'er--Rao bound of every parameter is bounded by a user-specified tolerance.
For the virtual model, this window is an interval of ratios $\rho = \gamma/\sigma$, so computing the IML estimate reduces to a one-dimensional search over $\rho$.
We apply AdaptivePnP to two standard PnP methods, yielding algorithms that require neither a tuned $\gamma$ nor a known $\sigma$.
On image deblurring and super-resolution, AdaptivePnP matches or outperforms PnP baselines tuned with the true $\sigma$.
Code is available at \url{https://github.com/hdubasullivan/AdaptivePnP}.
\end{abstract}


\section{Introduction} \label{sec:intro}

\IEEEPARstart{P}{lug-and-play} (PnP) methods are a standard approach for solving inverse imaging problems such as deblurring, super-resolution (SR), and tomographic reconstruction~\cite{venkatakrishnan_2013_plugandplay, zhang_2021_dpir, hurault_2022_gspnp, kamilov_2023_plugandplay}.
Rather than encoding the image prior as an explicit regularizer, PnP methods represent it implicitly through an image denoiser, which allows powerful learned priors to be combined with an explicit model of the measurement physics.
 
The quality and convergence of PnP reconstructions depend strongly on the step size $\gamma$ and the measurement noise level $\sigma$, which control the denoising strength and data-fitting step-size~\cite{wohlberg_2017_admm, xu_2017_relaxedadmm}.
In practice, $\gamma$ is fixed or annealed along a schedule tuned by grid search over validation images, and the tuning must be repeated whenever the problem setting changes~\cite{zhang_2021_dpir, hurault_2022_gspnp}.
Meanwhile, $\sigma$ is assumed to be known or estimated separately using the measurements~\cite{liu_2013_singleimage}.

Several approaches select one or both of these parameters automatically.
Learning-based methods use task-specific data to train a selection rule in advance, but they must be retrained whenever the imaging setting changes~\cite{zhang_2018_istaNet,yang_2016_admmNet,wei_2022_tfpnp,li_2025_curvpnp, liu_2025_plugandplay}.
Error-based criteria select the regularization strength from a scalar measure of fit or risk such as the discrepancy principle~\cite{morozov_1968_discrepancy}, Stein's unbiased risk estimate~\cite{stein_1981_sure}, generalized cross-validation~\cite{golub_1979_gcv}, or the L-curve~\cite{hansen_1993_use}.
Adaptive penalty and step-size rules tune an algorithmic parameter along the trajectory to speed up convergence~\cite{he_2000, wohlberg_2017_admm, xu_2017_adaptivepenalty, barzilai_1988, malitsky_2024_adaptivepgm}.
However, none of these approaches estimate $\sigma$ and $\gamma$ jointly.

Statistical model-based approaches come closer to joint estimation by fitting the parameters to a probabilistic model of an observed quantity.
Approximate message passing (AMP) methods use a state-evolution prediction of the effective error variance~\cite{metzler_2016_denoising, karuppasamy_2024_distributed, kamilov_2012_approximate, fletcher_2017_learning}; however, AMP methods modify the underlying PnP iteration and assume that the forward model is a large i.i.d.\ sub-Gaussian matrix, often causing divergence when applied to convolutional forward models~\cite{vila_2015_adaptive, rangan_2019_convergence}.
Residual whiteness methods enforce whiteness of the measurement residual, typically through its normalized autocorrelation, but these methods determine only the ratio $\rho = \gamma/\sigma$~\cite{almeida_2013_parameter, lanza_2020_whiteness, pragliola_2023_whiteness, wang_2023_hyperspectral}.
Empirical Bayes methods maximize the marginal likelihood of the measurements, which yields a closed-form joint estimate of $\sigma$ and $\gamma$ under a Gaussian prior~\cite{galatsanos_1992_methods, mackay_1992_bayesian, molina_1999_bayesian}. 
For general image priors, however, the marginal likelihood is intractable; recent methods address this with Langevin sampling, at a substantial increase in computational cost~\cite{vidal_2020_maximum, mbakam_2024_marginal, kemajoumbakam_2025_empiricalbayes}.

In this paper, we propose AdaptivePnP, a PnP framework that jointly estimates $\sigma$ and $\gamma$ at every iteration from the data-fitting residual.
Like empirical Bayes, AdaptivePnP fits the parameters to a likelihood; however, the likelihood is that of the residual rather than of the measurements.
Applying the PnP virtual model~\cite[Ch.~10]{bouman_2022_foundations} to the data-fitting proximal map, we show that the residual is zero-mean Gaussian with covariance parameterized by $\sigma$ and $\gamma$, which contribute spectrally distinct components to the covariance.
Consequently, we can jointly estimate $\sigma$ and $\gamma$ and the estimator requires no task-specific training, no modification of the PnP iteration, and no explicit prior or sampling procedure.

\begin{figure*}
	\centering
	\includegraphics[width=0.9\linewidth]{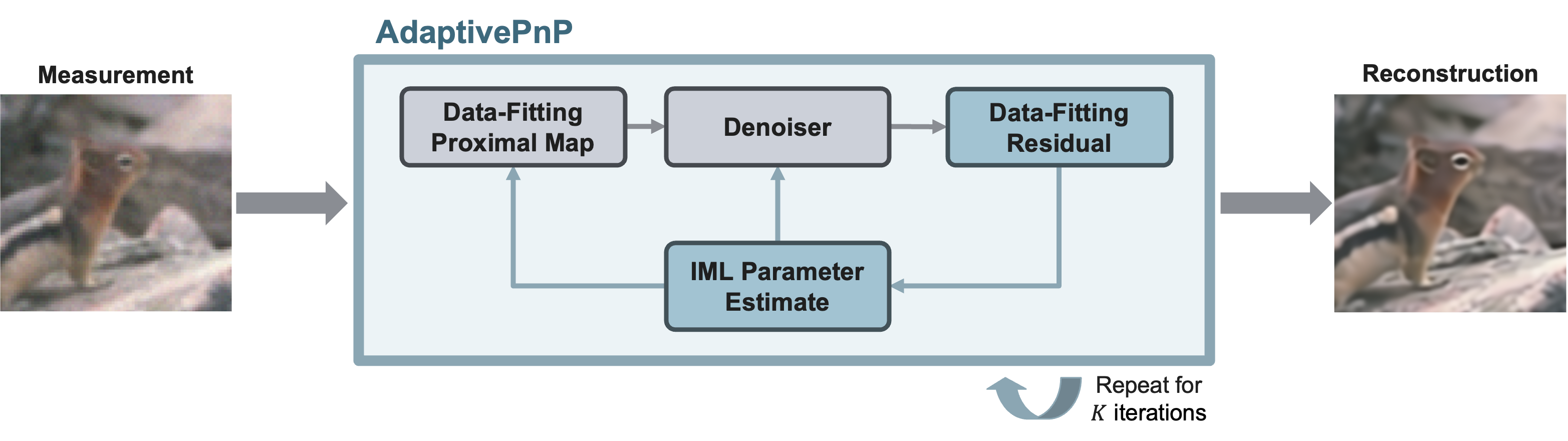}
	\caption{Overview of the proposed AdaptivePnP framework. Standard PnP methods alternate a data-fitting proximal map with an image denoiser (gray boxes); AdaptivePnP adds a parameter estimation step in each iteration (blue boxes). Namely, AdaptivePnP applies the identifiable maximum likelihood (IML) estimator to the data-fitting residual, resulting in a joint estimate of the noise level $\sigma$ and the step size $\gamma$. The estimates are fed back to both the denoiser and the data-fitting proximal map, so neither a tuned step size nor a known noise level is required. }
	\label{fig:pipeline}
\end{figure*}

However, when one parameter is much larger than the other, the residual carries little information about the smaller one, and the resulting maximum likelihood (ML) estimates can degrade reconstruction quality.
We therefore introduce the identifiable maximum likelihood (IML) estimator, which restricts the ML estimator to an identifiable window $\W_\tau$ where the normalized Cram\'er--Rao bound of both parameters is below a user-specified tolerance $\tau$.
Unlike identifiability analysis, which diagnoses poorly estimable parameters~\cite{jacquez_1985_numerical, miao_2011_identifiability, raue_2009_structural, gutenkunst_2007_universally, chis_2016_relationship,burth_1999_subset, cintron-arias_2020_sensitivity}, the IML estimator is a constrained ML estimator~\cite{aitchison_1958_maximumlikelihood, gorman_1990_lower} that restricts the estimate to values the data can resolve within a prescribed error tolerance.
Figure~\ref{fig:pipeline} shows the resulting proposed AdaptivePnP pipeline.

Our main contributions are:
\begin{itemize}
    \item A method for adaptive parameter estimation in PnP reconstruction that results in fast and accurate convergence;
    \item A novel identifiable ML estimator that eliminates parameter values the data cannot estimate within a prescribed error tolerance;
    \item A virtual model of the data-fitting proximal map that facilitates analysis and interpretation of PnP algorithms; and
    \item An efficient implementation for convolutional forward models, with experiments on deblurring and SR showing that AdaptivePnP matches or outperforms PnP baselines tuned with known $\sigma$.
\end{itemize}

\section{AdaptivePnP Framework}\label{sec:adaptivepnp}

Our goal is to recover an unknown image $x \in \R^N$ from observed measurements $y \in \R^M$ such that
\begin{equation}\label{eq:forward}
	y=Ax+\sigma w,
\end{equation}
where $A\in \R^{M \times N}$ is a linear forward operator, $w \sim \N(0,I)$ is additive white Gaussian noise (AWGN), and $\sigma>0$ is the standard deviation of the measurement noise.

The maximum a posteriori (MAP) estimate of $x$ is 
\begin{equation}\label{eq:map}
	\hat{x} = \arg \min_x\left\{f_\sigma(x) + g(x)\right\},
\end{equation}
where $f_\sigma(x) = -\log p(y\mid x)$ is the negative log-likelihood and $g(x)= -\log p(x)$ is the negative log-prior.
Under the measurement model in~\eqref{eq:forward},
\begin{equation}\label{eq:f-nll}
	f_\sigma(x)=\frac{1}{2\sigma^2}\|y-Ax\|^2.
\end{equation}
The proximal map of $f_\sigma(x)$ with step size $\gamma > 0$ is given by
\begin{equation}\label{eq:data-fitting-prox}
    F_{\sigma,\gamma}(v;y) = \arg\min_x \left\{ f_\sigma(x) + \frac{1}{2\gamma^2}\|x-v\|^2 \right\}.
\end{equation} 

Instead of using $g$ explicitly, PnP methods alternate the data-fitting proximal map $F_{\sigma,\gamma}$ with an image denoiser $D_\gamma$ for AWGN of standard deviation $\gamma$.
The denoiser $D_\gamma$ depends only on $\gamma$, while $F_{\sigma,\gamma}$ depends only on the ratio $\rho = \gamma/\sigma$.
As discussed in Section~\ref{sec:intro}, choosing these parameters is a key question when using PnP methods.

Our proposed AdaptivePnP framework removes the need to choose these parameters in advance by instead estimating them from the iterates themselves.
At iteration $k$, we form the data-fitting residual $z_k = y - Av_k$ from the current denoised estimate $v_k$ and estimate $(\hat\sigma_k, \hat\gamma_k)$ from $z_k$.

Since these estimates are fed back into the next iteration, AdaptivePnP works best when $\sigma$ and $\gamma$ are identifiable from the residual.
When one parameter dominates the other, the residual carries little information about the smaller one, and we show in Section~\ref{sec:results-iml} that the resulting ML estimates can degrade reconstruction quality.
We therefore use the IML estimator, which restricts the ML estimator to parameter values that the residual can resolve to a prescribed relative tolerance $\tau>0$.
Section~\ref{sec:IML} specifies the model of $z_k$ and defines the IML estimator.

We apply AdaptivePnP to the proximal gradient method (PGM) and the alternating direction method of multipliers (ADMM), resulting in the AdaptivePnP-PGM and AdaptivePnP-ADMM methods shown in Algorithms~\ref{alg:AdaptivePnP-PGM} and~\ref{alg:AdaptivePnP-ADMM}.
We re-estimate the parameters immediately after each denoising step using the IML estimator, i.e.,
\begin{equation}\label{eq:iml-call}
	(\hat\sigma_k^2, \hat\gamma_k^2) = \text{IML}(z_k, \mathcal{P}),
\end{equation} 
where $\mathcal{P} = \text{Precompute}(A,\tau)$ collects the quantities IML requires from the forward model and the tolerance, computed once before reconstruction.
The only parameters that must be set are the number of PnP iterations $K$ and the tolerance $\tau$.
We observe empirically that reconstruction quality is insensitive to $\tau$ once it is large enough, so we fix $\tau = 0.05$.

\begin{figure*}
\begin{minipage}[t]{0.48\textwidth}
\begin{algorithm}[H]
\caption{AdaptivePnP-PGM}\label{alg:AdaptivePnP-PGM}
\begin{algorithmic}[1]
\renewcommand{\algorithmicrequire}{\textbf{Input:}}
\Require $y$, $A$, $K$, $\tau$
\vspace{1mm}
\State $\mathcal{P} \gets \text{Precompute}(A, \tau)$ \Comment{Alg.~\ref{alg:precompute}}
\vspace{1mm}
\State $x_0 \gets A^\T y$
\HState{$z_0 \gets y - Ax_0$}
\HState{$(\hat{\sigma}_{0}^2, \hat{\gamma}_{0}^2) \gets \text{IML}(z_0, \mathcal{P})$ \Comment{Alg.~\ref{alg:iml}}}
\vspace{0.5mm}
\For{$k=0,\dots,K-1$}
    \State $v_{k+1} \gets D_{\hat{\gamma}_{k}}(x_k)$ \Comment{denoising}
    \HState{$z_{k+1} \gets y - Av_{k+1}$}
    \HState{$(\hat{\sigma}_{k+1}^2, \hat{\gamma}_{k+1}^2) \gets \text{IML}(z_{k+1}, \mathcal{P})$ \Comment{Alg.~\ref{alg:iml}}}
    \State $x_{k+1} \gets F_{\hat{\sigma}_{k+1},\hat{\gamma}_{k+1}}(v_{k+1};\, y)$ \Comment{data fitting}
\EndFor
\State \Return $x_K$
\end{algorithmic}
\end{algorithm}
\end{minipage}\hfill
\begin{minipage}[t]{0.48\textwidth}
\begin{algorithm}[H]
\caption{AdaptivePnP-ADMM}\label{alg:AdaptivePnP-ADMM}
\begin{algorithmic}[1]
\renewcommand{\algorithmicrequire}{\textbf{Input:}}
\Require $y$, $A$, $K$, $\tau$
\vspace{1mm}
\State $\mathcal{P} \gets \text{Precompute}(A, \tau)$ \Comment{Alg.~\ref{alg:precompute}}
\vspace{1mm}
\State $v_0 \gets A^\T y$, \; $u_0 \gets 0$
\HState{$z_0 \gets y - Av_0$}
\HState{$(\hat{\sigma}_{0}^2, \hat{\gamma}_{0}^2) \gets \text{IML}(z_0, \mathcal{P})$ \Comment{Alg.~\ref{alg:iml}}}
\vspace{0.5mm}
\For{$k=0,\dots,K-1$}
    \State $x_{k+1} \gets F_{\hat{\sigma}_k,\hat{\gamma}_k}(v_k - \hat{\gamma}_{k}^2 u_k;\, y)$ \Comment{data fitting}
    \State $v_{k+1} \gets D_{\hat{\gamma}_k}(x_{k+1} + \hat{\gamma}_{k}^2 u_k)$ \Comment{denoising}
    \HState{$z_{k+1} \gets y - Av_{k+1}$}
    \HState{$(\hat{\sigma}_{k+1}^2, \hat{\gamma}_{k+1}^2) \gets \text{IML}(z_{k+1}, \mathcal{P})$ \Comment{Alg.~\ref{alg:iml}}}
    \State $u_{k+1} \gets u_k + \hat{\gamma}_{k}^{-2}(x_{k+1} - v_{k+1})$ \Comment{dual update}
\EndFor
\State \Return $v_K$
\end{algorithmic}
\end{algorithm}
\end{minipage}

\vspace{2mm}
{\footnotesize \noindent Shaded lines are the steps added by AdaptivePnP, which compute the residual and re-estimate the parameters. The remaining lines are standard PnP-PGM and PnP-ADMM with $(\hat\sigma_k,\hat\gamma_k)$ in place of fixed $(\sigma,\gamma)$.\par}
\end{figure*}

\section{Identifiable Maximum Likelihood Estimator}\label{sec:IML}

\subsection{Definition of IML Estimator}\label{sec:IML-general}

Consider estimating a parameter vector $\theta \in \Theta \subseteq (0,\infty)^d$ from a random vector $\tilde Z \in \R^M$ with likelihood $p_\theta(\tilde z)$.
The standard ML estimator selects the parameter values that maximize the likelihood of the observed data $\tilde{z}$.
However, $\tilde{z}$ may barely depend on one of the parameters, in which case many different values of that parameter explain the data nearly equally well.
The Cram\'er--Rao bound quantifies how well the data can determine each parameter of the vector $\theta$.
Our proposed IML estimator uses this bound to eliminate values of $\theta$ for which some parameter $\theta_i$ cannot be estimated within a specified error.

Let $\I(\theta) = -\E[\nabla_\theta^2 \log p_\theta(\tilde Z)]$ be the Fisher information matrix of $p_\theta$, which  is nonsingular on $\Theta$.
We define the normalized Cram\'er--Rao bound of parameter $\theta_i$ as
\begin{equation}\label{eq:v_i}
  V_i(\theta) = \frac{[\I(\theta)^{-1}]_{i,i}}{\theta_i^2}.
\end{equation}
Then, by the Cram\'er--Rao bound, $\sqrt{V_i(\theta)}$ is a lower bound on the normalized root mean squared error (NRMSE) of any unbiased estimate of $\theta_i$.

For a tolerance $\tau > 0$, we define the \textit{identifiable window} $\W_\tau$ as
\begin{equation}\label{eq:general-window}
  \W_\tau=\left\{\theta \in \Theta : V_i (\theta) \leq \tau^2 \text{ for all } i \right\}.
\end{equation}
Outside this window, the normalized Cram\'er--Rao bound exceeds the tolerance for at least one parameter, so no unbiased estimator can estimate that parameter within NRMSE $\tau$.
We therefore say that $\theta$ is \textit{identifiable} to tolerance $\tau$ when $\theta \in \W_\tau$.

We then define the IML estimator as the standard ML estimator constrained to this identifiable window,
\begin{equation}\label{eq:iml}
  \hat\theta_{\mathrm{IML}} = \argmax_{\theta\in\W_\tau}p_\theta(\tilde z),
\end{equation}
provided $\W_\tau$ is nonempty.
When the standard ML estimate lies in $\W_\tau$, the two estimates coincide; otherwise, the IML estimator returns the most likely parameter among those the data can resolve to the stated tolerance.

\subsection{Virtual Model}\label{sec:model}

In PnP methods, the prior proximal map is interpreted as a MAP denoiser by introducing a \emph{virtual model} in which the denoiser input is the true image plus AWGN of standard deviation $\gamma$~\cite[Ch.~10]{bouman_2022_foundations}.
This virtual model is not assumed to describe how the data were generated.
Instead, it is a probabilistic interpretation of an algorithmic step that makes the step's statistical meaning explicit.

Here, we apply the idea of the virtual model to the data-fitting proximal map of~\eqref{eq:data-fitting-prox}.  
From a Bayesian viewpoint, $f_\sigma(x)$ encodes the data-fitting loss, and the quadratic term $\frac{1}{2\gamma^2}\|x-v\|^2$ can be interpreted as the negative log of a Gaussian prior $\N(v,\gamma^2 I)$ on $x$.  
Hence, the minimization in~\eqref{eq:data-fitting-prox} gives $F_{\sigma,\gamma}(v;y)$ as the MAP estimate of $x$ under this prior and the data-fitting term from~\eqref{eq:f-nll}~\cite{dubasullivan_2026_PPM}.

Stated explicitly, the virtual model is
\begin{equation}\label{eq:x}
  X = V + \gamma\,\mathcal{E}, \qquad \mathcal{E} \sim \N(0,I),
\end{equation}
where $V \sim p(v)$ is a random vector drawn from the image prior. 
Then the measurement under the forward model in~\eqref{eq:forward} is
\begin{equation}\label{eq:y}
  Y = AX + \sigma W, \qquad W \sim \N(0,I).
\end{equation}
As noted above, under this virtual model, $F_{\sigma,\gamma}(v;y)$ is the MAP estimate of $X$ given $V=v$ and $Y=y$.  

To better connect this model with neural networks, we note that these MAP estimates are MMSE estimates.  That is, since the posterior of $X | v, y$ is Gaussian, the mode and mean coincide, i.e.,
\begin{equation}
    F_{\sigma,\gamma}(v;y) = \E[X \mid V=v, Y=y].
\end{equation}
Likewise, \eqref{eq:x} is exactly the AWGN denoising problem for standard deviation $\gamma$, so the MMSE denoiser is given by
\begin{equation}
    D_\gamma(x) = \E[V \mid X=x].
\end{equation}
Thus, the PnP iterations in Algorithms~1 and 2 alternate MMSE estimates of the two latent variables $X$ and $V$, with the measurement $Y=y$ held fixed.

This virtual model allows us to describe the likelihood $p_\theta$ of \eqref{eq:iml}, where $\theta = (\sigma^2, \gamma^2)$.  
Under the virtual model, define the data-fitting residual as $Z = Y - AV$.
Substituting \eqref{eq:x} and \eqref{eq:y} and using linearity of $A$, we have
\begin{equation}\label{eq:residual}
  Z = \gamma A\mathcal{E} + \sigma W \sim \N\!\left(0,\, \gamma^2 AA^\T + \sigma^2 I\right).
\end{equation}
Thus, the residual is the sum of two statistically distinct parts: a reconstruction error $\gamma A\mathcal{E}$ that is shaped by the forward operator, and measurement noise $\sigma W$ that is white.
This is what makes $\sigma$ and $\gamma$ separately identifiable as long as neither part overwhelms the other.

To make this explicit, let $AA^\T = E\Lambda E^\T$ be the eigendecomposition of $AA^\T$, where $E$ is orthogonal and $\Lambda$ is diagonal with nonnegative entries $\lambda_i$.
We assume the $\lambda_i$ are not all equal; otherwise the residual would be white and the two parameters could not be separated.
Lemma~\ref{lem:fisher} shows that this is exactly the condition for the Fisher information matrix to be nonsingular.
Rotating by $E$ decorrelates the residual, so $\tilde Z = E^\T Z$ has independent components with variances $\gamma^2\lambda_i + \sigma^2$.
Factoring out $\sigma^2$ and writing $\rho = \gamma/\sigma$, we have
\begin{equation}\label{eq:WhitenedModel}
  \tilde Z \sim \N(0, \sigma^2 C_\rho), \qquad C_\rho = \rho^2\Lambda + I.
\end{equation}
The covariance thus separates into an overall scale $\sigma^2$ and a spectral shape $C_\rho$ that depends only on the ratio $\rho$.

\subsection{IML Estimator for Virtual Model}\label{sec:IML-application}

We now provide the complete algorithm for computing the IML estimate of the PnP parameters $\theta = (\sigma^2,\gamma^2)$ from the rotated residual $\tilde Z \sim \N(0,\sigma^2 C_\rho)$.
All derivations and proofs are included in Appendix~\ref{app:derivations}.

The identifiable window $\W_\tau$ of \eqref{eq:general-window} is determined by the normalized Cram\'er--Rao bounds of $\gamma^2$ and $\sigma^2$; these bounds are given by
\begin{align}
	\label{eq:Vgamma}
	V_\gamma(\rho; \Lambda) &= \frac{2\Tr(C_\rho^{-2})}{\rho^4\Delta(\rho)},\\
	\label{eq:Vsigma}
	V_\sigma(\rho; \Lambda) &= \frac{2\Tr(\Lambda^2C_\rho^{-2})}{\Delta(\rho)},
\end{align}
where
\begin{equation}\label{eq:Delta}
	\Delta(\rho) = \Tr(C_\rho^{-2})\Tr(\Lambda^2 C_\rho^{-2}) - \bigl[\Tr(\Lambda C_\rho^{-2})\bigr]^2.
\end{equation}
As shown in Lemma~\ref{lem:fisher}, both bounds depend on $\theta$ only through the ratio $\rho$, so $\W_\tau$ constrains only $\rho$.  This is consistent with the earlier observation that $\sigma$ and $\gamma$ are separately identifiable as long as neither overwhelms the other. 

Theorem~\ref{thm:mono} shows that $V_\gamma$ is decreasing and $V_\sigma $ is increasing in $\rho$, so each constraint bounds $\rho$ on one side. 
Thus, given $\tau > 0$, $\W_\tau$ is the set of parameter pairs $(\sigma^2, \gamma^2)$ such that
\begin{equation}\label{eq:rho-interval}
	\rho_{\inf} \leq \rho \leq \rho_{\sup},
\end{equation}
where $\rho_{\inf}$ solves $V_\gamma(\rho;\Lambda) = \tau^2$ and $\rho_{\sup}$ solves $V_\sigma(\rho;\Lambda) = \tau^2$, with $\rho_{\sup} = \infty$ if $V_\sigma(\rho;\Lambda) < \tau^2$ for all $\rho$.
Since both functions are monotone, each root can be found by bisection.
Algorithm~\ref{alg:precompute} summarizes the computation of $\rho_{\inf}$ and $\rho_{\sup}$.
We assume that $\tau$ is chosen to be large enough that this interval is nonempty.
Corollary~\ref{cor:tau-star} shows that the interval is nonempty if and only if $\tau \geq \tau_\star$, where $\tau_\star$ is the common value of the two bounds at the $\rho$ where they cross.

\begin{algorithm}[t]
	\caption{\text{Precompute}$(A,\tau)$}\label{alg:precompute}
	\begin{algorithmic}[1]
		\renewcommand{\algorithmicrequire}{\textbf{Input:}}
		\Require forward operator $A$, tolerance $\tau$
		\vspace{1mm}
		\State $E, \Lambda \gets$ eigendecomposition of $AA^\T$
		\State $\rho_{\inf} \gets$ solution of $V_\gamma(\rho;\Lambda) = \tau^2$ \Comment{\eqref{eq:Vgamma}}
		\State $\rho_{\sup} \gets$ solution of $V_\sigma(\rho;\Lambda) = \tau^2$, or $\infty$ if none \Comment{\eqref{eq:Vsigma}}
		\State \Return $\mathcal{P} = (E, \Lambda, \rho_{\inf}, \rho_{\sup})$
	\end{algorithmic}
\end{algorithm}

Algorithm~\ref{alg:iml} computes the IML estimate from a residual $z$.
Since $\W_\tau$ constrains only $\rho$, the likelihood is first maximized over $\sigma^2$ for fixed $\rho$, which gives 
\begin{equation}\label{eq:sig_hat}
    \hat\sigma^2(\rho) = \frac{1}{M}\tilde z^\T C_\rho^{-1}\tilde z.
\end{equation}
Substituting this into the negative log-likelihood leaves the profiled negative log-likelihood of $\rho$,
\begin{equation}
	\mathcal{L}(\rho; \tilde{z}, \Lambda) = M \log\left( \tilde{z}^\T C_\rho^{-1} \tilde{z} \right) + \log \det C_\rho.
	\label{eq:Profile}
\end{equation}
The IML estimate of $\rho$ minimizes $\mathcal{L}$ over $[\rho_{\inf}, \rho_{\sup}]$, and Theorem~\ref{thm:exist} shows that this minimizer exists.
Since $\Lambda$ is diagonal, every trace and quadratic form is a sum over the $M$ eigenvalues, so each evaluation costs $O(M)$.

\begin{algorithm}[t]
\caption{\text{IML}$(z, \mathcal{P})$}\label{alg:iml}
\begin{algorithmic}[1]
\renewcommand{\algorithmicrequire}{\textbf{Input:}}
\Require residual $z \in \R^M$, IML parameters $\mathcal{P}$
\vspace{1mm}
\State $(E, \Lambda, \rho_{\inf}, \rho_{\sup}) \gets \mathcal{P}$ \Comment{Alg.~\ref{alg:precompute}}
\State $\tilde z \gets E^\T z$
\State $\hat\rho \gets \arg\min_{\rho_{\inf} \le \rho \le \rho_{\sup}} \mathcal{L}(\rho; \tilde z,\Lambda)$ \Comment{\eqref{eq:Profile}}
\State $C_{\hat\rho} \gets \hat\rho^2 \Lambda + I$
\State $\hat\sigma^2 \gets \frac{1}{M}\, \tilde z^\T C_{\hat\rho}^{-1} \tilde z$
\State $\hat\gamma^2 \gets \hat\rho^2 \hat\sigma^2$
\State \Return $(\hat\sigma^2, \hat\gamma^2)$
\end{algorithmic}
\end{algorithm}

\subsection{Implementation for Convolutional Forward Models} \label{sec:conv-implementation}

The IML estimator applies to any forward operator for which the eigendecomposition of $AA^\T$ is available.
In this work, we restrict our attention to convolutional forward models, specifically image deblurring and SR.

For deblurring, $A$ is circular convolution with blur kernel $k$, so that $Ax = k \ast x$.
Let $\mathcal{F}$ denote the discrete Fourier transform (DFT) matrix and $\hat k = \mathcal{F} k_{\mathrm{pad}}$, where $k_{\mathrm{pad}}$ is the kernel zero-padded to the image dimensions.
Circular convolution is diagonalized by the DFT, so $AA^\T = \frac{1}{M}\mathcal{F}^*\diag\{|\hat k_i|^2\}\mathcal{F}$ and the eigenvalues of $AA^\T$ are the power spectrum of the kernel, i.e.,
\begin{equation}\label{eq:deblur-eig}
  \lambda_i = |\hat k_i|^2.
\end{equation}
The rotation $\tilde z = E^\T z$ is the unitary DFT, i.e., $\tilde z = \frac{1}{\sqrt{M}}\mathcal{F} z$.\footnote{Since $\mathcal{F}$ is complex, $E = M^{-1/2}\mathcal{F}^*$ is unitary rather than orthogonal. A real orthogonal eigenbasis of $AA^\T$ is obtained from the real and imaginary parts of the DFT basis, and every quantity in Section~\ref{sec:IML-application} depends on $\tilde z$ only through $|\tilde z_i|^2$, so we compute with the complex DFT and replace $\tilde z^\T$ with $\tilde z^{\mathsf H}$.}

For SR by an integer factor $L > 1$, the forward model is $Ax = S(k \ast x)$, where $S$ subsamples the blurred image by a factor of $L$ in each spatial dimension.
Under circular boundary conditions, $AA^\T$ is circulant on the low-resolution grid and is therefore diagonalized by the low-resolution DFT~\cite{zhao_2016_fast}.
Its eigenvalues are the aliased power spectrum
\begin{equation}\label{eq:aliased_spectrum}
  \lambda_i = \frac{1}{L^2}\sum_{j \in \mathcal{A}_i} |\hat k_j|^2,
\end{equation}
where $\mathcal{A}_i$ is the set of $L^2$ high-resolution frequencies that alias onto low-resolution frequency $i$, and the rotation $E^\T z$ is the unitary DFT on the low-resolution grid.

\section{Experimental Setup}\label{sec:setup}

\begin{figure}
	\centering
	\includegraphics[width=0.85\linewidth]{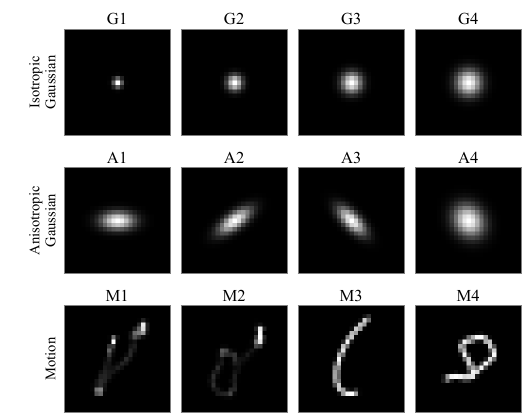}
	\caption{The twelve $25\times25$ blur kernels used in experiments. }
	\label{fig:kernels}
\end{figure}

\begin{table*} 
	\caption{Mean PSNR / SSIM of AdaptivePnP and tuned PnP reconstructions, averaged over the CBSD68 dataset with four kernels in each family (G1--G4, A1--A4, M1--M4) and three noise levels (1\%, 3\%, and 5\%). Tuned PnP uses the true $\sigma$ and a step size selected by grid search, whereas AdaptivePnP uses neither.The best result within each algorithm and kernel family is bolded. }
	\label{tab:metric_comp}
	\centering
	\setlength{\tabcolsep}{5pt}
	\begin{tabular}{l c cc cc cc}
		\toprule
		\multicolumn{2}{l}{} & \multicolumn{2}{c}{Isotropic Gaussian} & \multicolumn{2}{c}{Anisotropic Gaussian} & \multicolumn{2}{c}{Motion} \\
		\cmidrule(lr){3-4}\cmidrule(lr){5-6}\cmidrule(lr){7-8}
		&   & Tuned PnP & AdaptivePnP & Tuned PnP & AdaptivePnP & Tuned PnP & AdaptivePnP \\
		\midrule
		\multirow{2}{*}{Deblurring}
		& PGM & 27.83 / 0.768 & \textbf{28.39} / \textbf{0.807} & 25.49 / 0.701 & \textbf{25.83} / \textbf{0.719} & 26.64 / 0.719 & \textbf{28.01} / \textbf{0.777} \\
		& ADMM & 28.27 / 0.793 & \textbf{28.32} / \textbf{0.808} & 25.77 / \textbf{0.722} & \textbf{25.79} / 0.719 & \textbf{28.02} / \textbf{0.800} & 27.96 / 0.782 \\
		\midrule
		\multirow{2}{*}{2$\times$ Super-Resolution}
		& PGM & 24.07 / 0.603 & \textbf{26.69} / \textbf{0.737} & 24.83 / 0.634 & \textbf{25.23} / \textbf{0.675} & 22.24 / 0.470 & \textbf{25.47} / \textbf{0.650} \\
		& ADMM & 23.88 / 0.623 & \textbf{26.55} / \textbf{0.737} & 25.13 / 0.671 & \textbf{25.28} / \textbf{0.685} & 23.03 / 0.578 & \textbf{25.79} / \textbf{0.701} \\
		\bottomrule
	\end{tabular}
	
\end{table*}

We evaluate AdaptivePnP on image deblurring and $2\times$ SR using the CBSD68 dataset~\cite{martin_2001_database}.
Comparisons with the tuned baselines use all 68 images; the ablations in Sections~\ref{sec:results-iml} and~\ref{sec:results-sigma} use only the first 20 images.
We generate measurements using the twelve $25\times25$ blur kernels shown in Figure~\ref{fig:kernels}, which are distributed with DPIR~\cite{zhang_2021_dpir}.
This set of kernels consists of four isotropic Gaussian kernels (G1--G4), four anisotropic Gaussian kernels (A1--A4), and four motion blur kernels (M1--M4).
Each kernel is normalized to unit sum and applied with circular boundary conditions, consistent with Section~\ref{sec:conv-implementation}.
For both forward models, we use measurement noise levels $\sigma\in\{0.01,0.03,0.05\}$ corresponding to $1\%$, $3\%$, and $5\%$ noise.

All methods use the same pretrained color DRUNet denoiser distributed with DPIR~\cite{zhang_2021_dpir}.
Additionally, we initialize with the back-projected measurement $A^{\T}y$ and run for $K=10$ iterations with no early stopping.
We report PSNR and SSIM computed on the RGB reconstruction against the ground truth image, excluding a border of 10 pixels.

For all AdaptivePnP results, we use $\tau=0.05$ based on a sensitivity analysis included in Section~\ref{sec:results-iml}.
We compute $\rho_{\inf}$ and $\rho_{\sup}$ by bracketing each root on a 400-point logarithmic grid over $\rho\in[10^{-6},10^{6}]$, and refine it with Brent's method as implemented in SciPy~\cite{brent2013algorithms, virtanen2020scipy}.
If $V_\sigma$ does not reach $\tau^2$ on this range, we set $\rho_{\sup} = 10^{6}$.
At each PnP iteration, the profiled negative log-likelihood from~\eqref{eq:Profile} is minimized over $[\rho_{\inf},\rho_{\sup}]$ with SciPy's bounded Brent minimizer.
On an NVIDIA A100 GPU, one call of Algorithm~\ref{alg:iml} on a $481\times321$ image takes approximately 6\,ms, compared with 12\,ms for one DRUNet evaluation.

The tuned PnP baselines are given the true measurement noise level $\sigma$ and a fixed step size $\gamma=\rho\sigma$.
The ratio $\rho$ is selected by grid search on a separate validation set of 20 images from DIV2K~\cite{agustsson_2017_ntire}, center-cropped to $256\times256$ and degraded with the same twelve kernels and three noise levels.
For each reconstruction algorithm and forward model, we evaluate a 25-point logarithmic grid over $\rho\in[10^{-2},10^{2}]$, and select the value with the highest mean PSNR pooled over all kernels, noise levels, and validation images.
This yields one validation-tuned $\rho$ per algorithm and forward model: $\rho=2.15$ for ADMM and $\rho=1.47$ for PGM in deblurring, and $\rho=4.64$ for ADMM and $\rho=3.16$ for PGM in $2\times$ SR.
We pool over kernels and noise levels rather than tuning each condition separately since this mirrors how fixed PnP parameters are most often used in practice.
Note that tuned PnP requires the true $\sigma$ and paired ground truth validation data to tune $\rho$, while AdaptivePnP doesn't need either of these.
 
\section{Experimental Results}\label{sec:results}

\subsection{Comparison With Tuned PnP Baselines}

Table~\ref{tab:metric_comp} reports the mean PSNR and SSIM of AdaptivePnP and tuned PnP after 10 iterations, averaged over the CBSD68 dataset with four kernels in each family and three noise levels.
For deblurring, AdaptivePnP-PGM outperforms tuned PnP-PGM for all three kernel families, with the largest improvement for motion blur.
AdaptivePnP-ADMM essentially matches the tuned ADMM baseline.
For SR, AdaptivePnP improves PSNR over tuned PnP by more than $2$\,dB for both isotropic Gaussian and motion blur kernels.

\begin{figure*} 
	\centering
	
	\subfloat[Deblurring]{
		\includegraphics[width=0.45\linewidth]{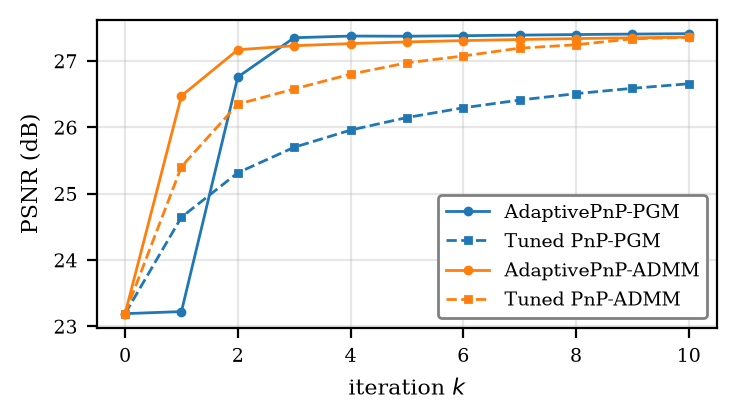}
		\hfill
		\includegraphics[width=0.45\linewidth]{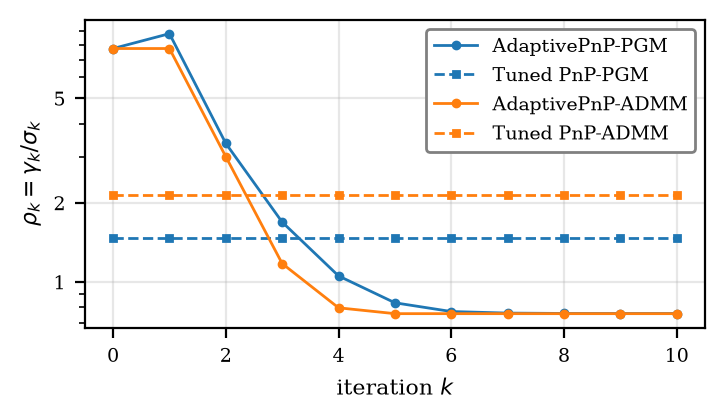}
	}
	
	\subfloat[2$\times$ Super-Resolution]{
		\includegraphics[width=0.45\linewidth]{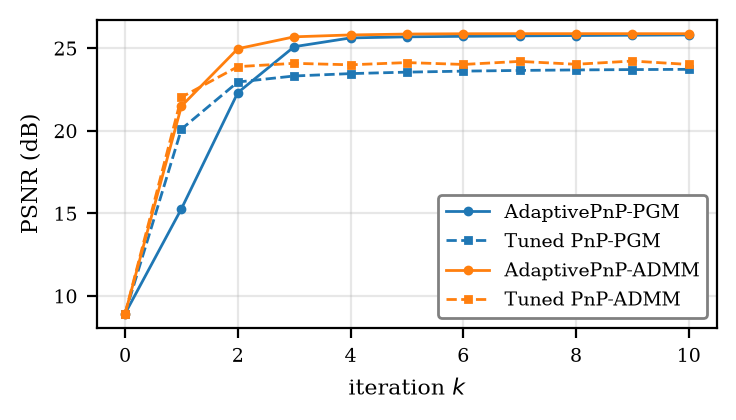}
		\hfill
		\includegraphics[width=0.45\linewidth]{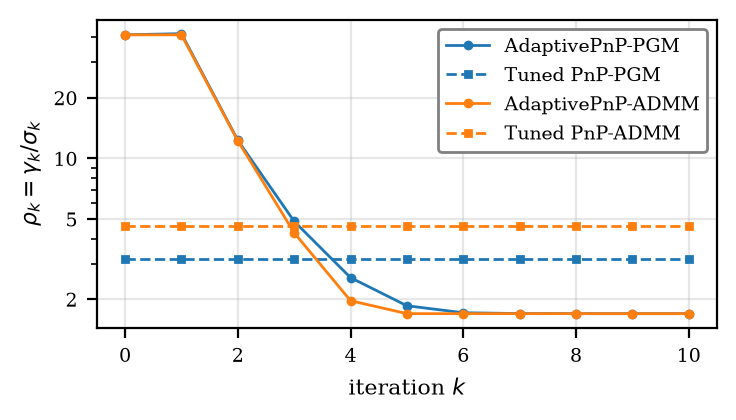}
	}
	
	\caption{Mean reconstruction PSNR and ratio $\rho_k=\gamma_k/\sigma_k$ of AdaptivePnP and tuned PnP over 10 iterations for (a) deblurring and (b) $2\times$ SR. Results are averaged over the CBSD68 dataset with all 12 kernels and three noise levels. AdaptivePnP rapidly adapts $\rho_k$ and reaches its final reconstruction quality within the first few iterations without requiring a validation-tuned parameter choice.}
	\label{fig:trajectory_comp}
\end{figure*}

Figure~\ref{fig:trajectory_comp} compares the PSNR and parameter ratio $\rho_k=\gamma_k/\sigma_k$ of AdaptivePnP and tuned PnP over 10 iterations, averaged over the CBSD68 dataset with all 12 kernels and three noise levels.
For both deblurring and $2\times$ SR, the AdaptivePnP estimate of $\rho_k$ decreases sharply over the first several iterations as the reconstruction PSNR increases; both stabilize within approximately five iterations.

\begin{figure*}
    \centering
    \includegraphics[width=\linewidth]{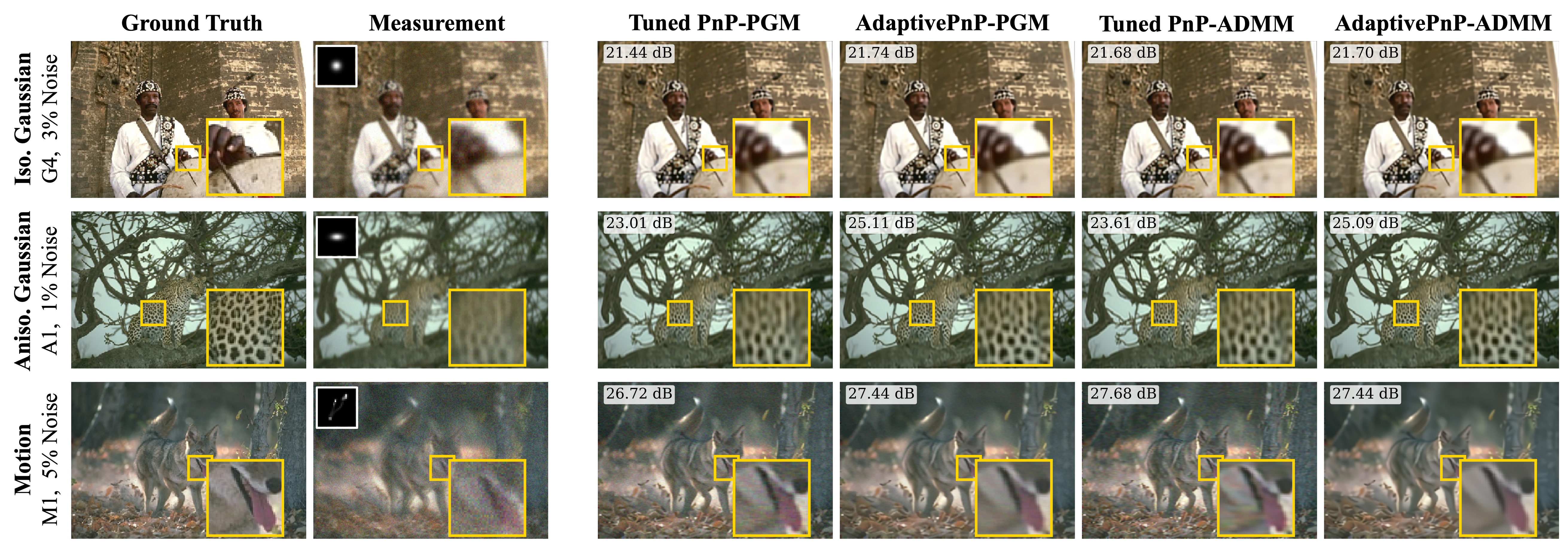}
    \caption{Deblurring reconstructions using AdaptivePnP and tuned PnP; rows show an isotropic Gaussian kernel (G4) with $3\%$ measurement noise, an anisotropic Gaussian kernel (A1) with $1\%$ noise, and a motion blur kernel (M1) with $5\%$ noise. The ground truth image and blurred noisy measurement are shown for reference, and the highlighted regions show magnified image details. AdaptivePnP produces visual quality comparable to or better than the tuned PnP baselines despite estimating both $\sigma$ and $\gamma$ directly from the reconstruction.}
    \label{fig:deblurring_comp}
\end{figure*}

\begin{figure*}
    \centering
    \includegraphics[width=\linewidth]{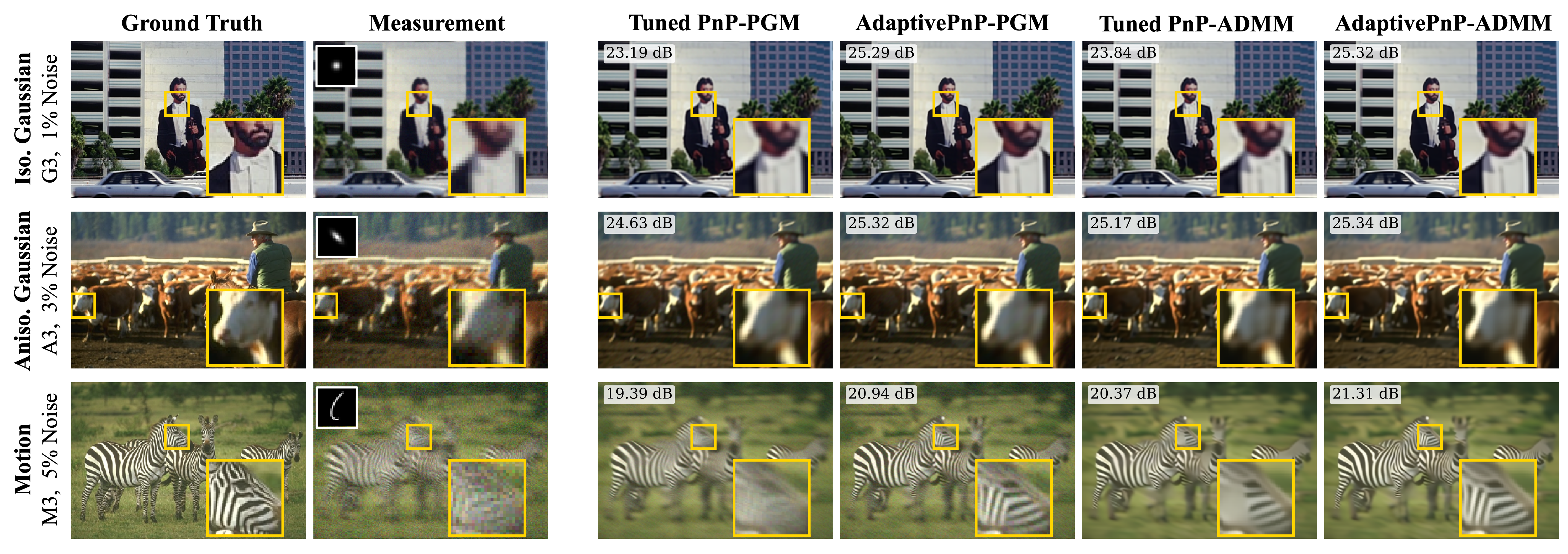}
    \caption{$2\times$ SR reconstructions using AdaptivePnP and tuned PnP; rows show an isotropic Gaussian kernel (G3) with $1\%$ measurement noise, an anisotropic Gaussian kernel (A3) with $3\%$ noise, and a motion blur kernel (M3) with $5\%$ noise. The ground truth image and low-resolution noisy measurement are shown for reference, and the highlighted regions show magnified image details. AdaptivePnP recovers visibly sharper image detail than the tuned PnP baselines, consistent with the PSNR and SSIM improvements in Table~\ref{tab:metric_comp}.}
    \label{fig:sr_comp}
\end{figure*}

Figures~\ref{fig:deblurring_comp} and~\ref{fig:sr_comp} compare deblurring and $2\times$ SR reconstructions from AdaptivePnP and tuned PnP for isotropic Gaussian, anisotropic Gaussian, and motion blur kernels at different measurement noise levels.
AdaptivePnP produces reconstructions comparable to or better than those obtained with the tuned PnP baselines, despite requiring neither validation-based tuning nor prior knowledge of the measurement noise level.

\subsection{Impact of IML Estimator}
\label{sec:results-iml}

\begin{figure*}
	\centering
	
	\subfloat[Deblurring]{
		\includegraphics[width=0.45\linewidth]{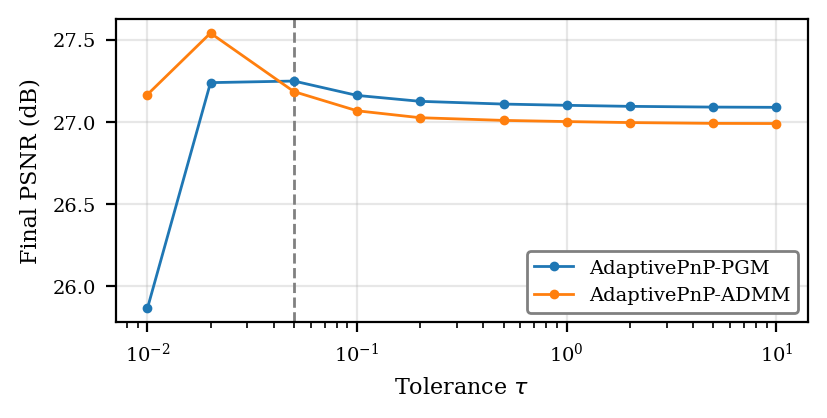}
		\label{fig:deblurring_tau_sweep}
	}
	\subfloat[2$\times$ Super-Resolution]{
		\includegraphics[width=0.45\linewidth]{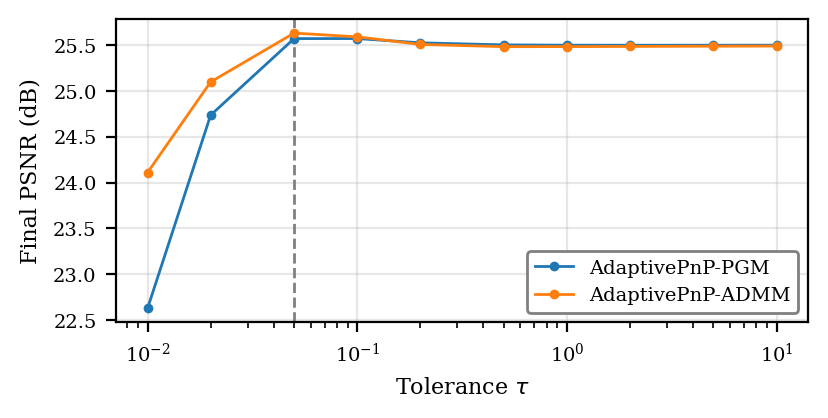}
		\label{fig:sr_tau_sweep}
	}
	\caption{Mean PSNR of AdaptivePnP after 10 iterations using varying tolerances $\tau$ for (a) deblurring and (b) $2\times$ SR. Results are averaged over the first 20 images from CBSD68 with all 12 kernels and three noise levels. Reconstruction quality is insensitive to the precise choice of the tolerance once it is sufficiently large; we use $\tau=0.05$ (dashed line) for all experiments.}
	\label{fig:tau_sweep}
\end{figure*}

Figure~\ref{fig:tau_sweep} shows the mean PSNR of AdaptivePnP after 10 iterations using varying tolerances $\tau$, averaged over the first 20 images from the CBSD68 dataset with all 12 kernels and three noise levels.
For $2\times$ SR, very small values of $\tau$ degrade reconstruction quality since the resulting identifiable window is overly restrictive, but performance is nearly constant for $\tau \geq 0.05$.
For deblurring, AdaptivePnP-ADMM is more sensitive to $\tau$ than AdaptivePnP-PGM, but the PSNR is within 0.5 dB of the optimal value for $\tau \geq 0.02$.
These results imply that the reconstruction is relatively insensitive to the precise choice of the tolerance once it is sufficiently large.
Thus, instead of tuning $\tau$ for optimal performance, we use $\tau=0.05$ for all experiments.

\begin{figure}
	\centering
	\subfloat[Deblurring]{
		\includegraphics[width=0.46\linewidth]{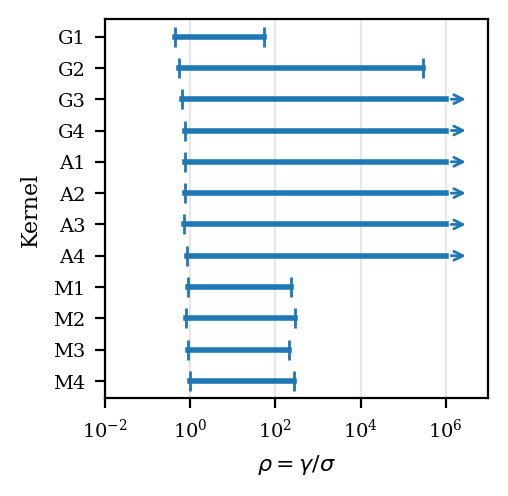}
		\label{fig:window_deblur}
	}
	\subfloat[2$\times$ Super-Resolution]{
		\includegraphics[width=0.46\linewidth]{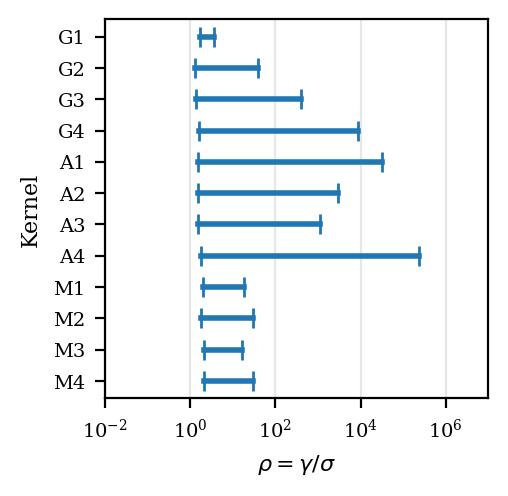}
		\label{fig:window_sr}
	}
	\caption{Identifiable windows $\W_\tau$ at $\tau = 0.05$ for all kernels under (a) deblurring and (b) $2\times$ SR. Horizontal intervals show $[\rho_{\inf}, \rho_{\sup}]$, with arrows indicating $\rho_{\sup} > 10^6$. All twelve windows are nonempty in both settings.}
	\label{fig:window}
\end{figure}

Figure~\ref{fig:window} shows the resulting identifiable window with $\tau = 0.05$ for each kernel.
The window is always nonempty; $\rho_{\inf}$ lies between 0.45 and 0.97 for deblurring and between 1.28 and 2.13 for $2\times$ SR, while $\rho_{\sup}$ exceeds $10^6$ for six of the deblurring kernels.
The threshold $\tau_\star$ of Corollary~\ref{cor:tau-star} ranges from 0.003 to 0.005 over the twelve kernels for deblurring and from 0.006 to 0.03 for $2\times$ SR.

\begin{figure*} 
	\centering
	
	\subfloat[Deblurring]{
		\includegraphics[width=0.45\linewidth]{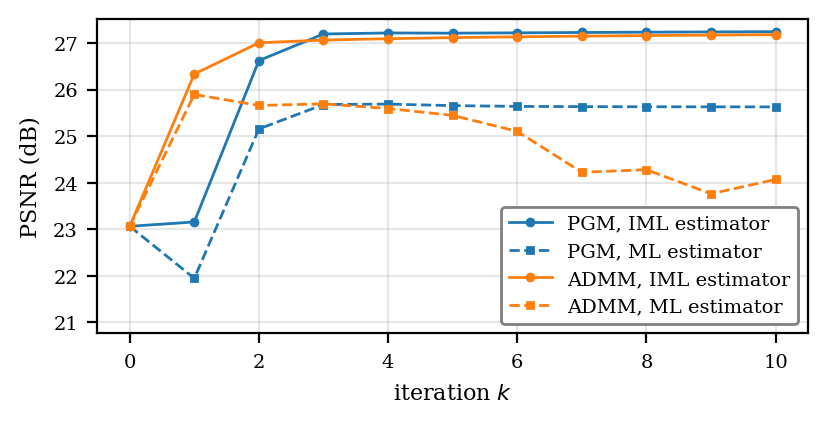}
	}
	\subfloat[2$\times$ Super-Resolution]{
		\includegraphics[width=0.45\linewidth]{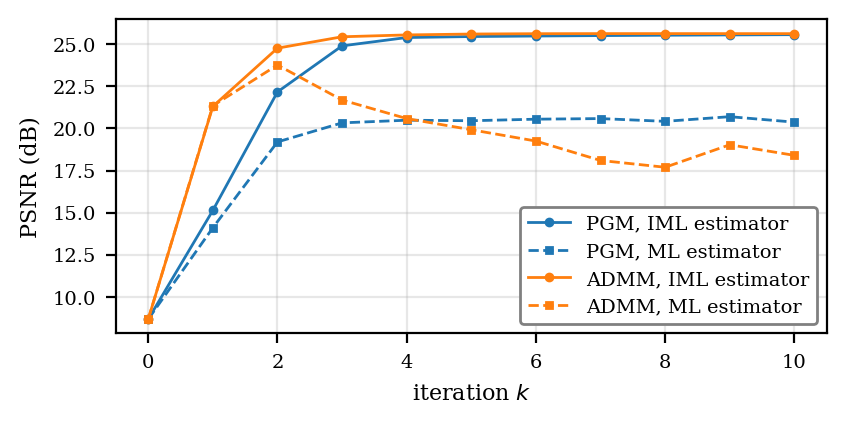}
	}
	
	\caption{Mean PSNR of AdaptivePnP over 10 iterations with the proposed IML estimator and the unconstrained ML estimator for (a) deblurring and (b) $2\times$ SR. Results are averaged over the first 20 images from CBSD68 with all 12 kernels and three noise levels. At each iteration, the mean PSNR excludes any runs that have diverged to a non-finite PSNR, which occurs only for the ML estimator. With IML, both algorithms converge within a few iterations, while using ML degrades quality over iterations.}
	\label{fig:IML_vs_ML}
\end{figure*}

Figure~\ref{fig:IML_vs_ML} compares the mean PSNR of AdaptivePnP over 10 iterations with the proposed IML estimator and the unconstrained ML estimator, averaged over the first 20 images in the CBSD68 dataset with all 12 kernels and three noise levels.
At each iteration, the mean PSNR excludes any runs that have diverged to a non-finite PSNR, which only occurs for the ML estimator.
With the IML estimator, both PGM and ADMM improve rapidly during the first few iterations before stabilizing at approximately 27\,dB for deblurring and 25\,dB for $2\times$ SR. 
In contrast, the unconstrained ML estimator exhibits substantially less stable behavior. 
For PGM, the mean PSNR plateaus at a lower value, while for ADMM, the mean PSNR initially increases but subsequently deteriorates, particularly for $2\times$ SR. 
These results demonstrate that constraining parameter estimates to the identifiable window improves both reconstruction stability and quality.

\subsection{Impact of Noise Level Estimation}
\label{sec:results-sigma}

\begin{figure*} 
	\centering
	
	\subfloat[Deblurring]{
		\includegraphics[width=0.45\linewidth]{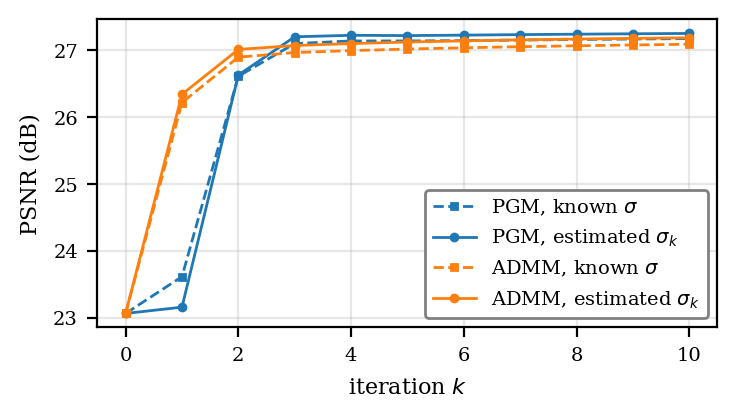}
	}
	\subfloat[2$\times$ Super-Resolution]{
		\includegraphics[width=0.45\linewidth]{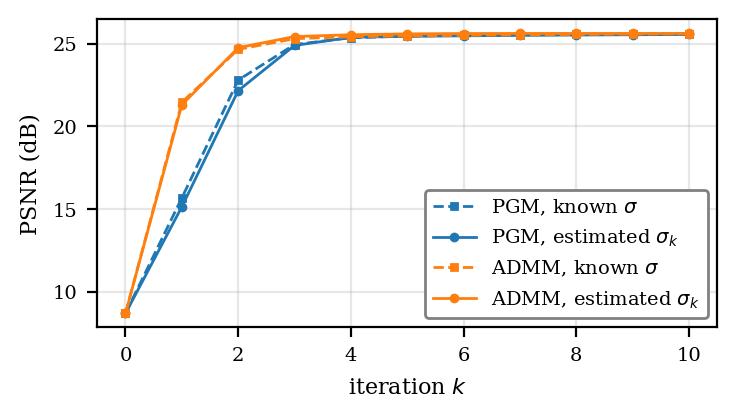}
	}
	\caption{Mean PSNR of AdaptivePnP over 10 iterations with estimated $\sigma_k$ and known $\sigma$ for (a) deblurring and (b) $2\times$ SR. Results are averaged over the first 20 images of CBSD68 with all 12 kernels and three noise levels; $\gamma$ is estimated by the IML estimator in both settings. AdaptivePnP removes the need for a known $\sigma$ at no cost in reconstruction quality.}
	\label{fig:sigma-ablation}
\end{figure*}

Figure~\ref{fig:sigma-ablation} compares the mean PSNR of AdaptivePnP over 10 iterations with estimated $\sigma_k$ and known $\sigma$, averaged over the first 20 images of CBSD68 with all 12 kernels and three noise levels.
In both settings, $\gamma$ is estimated at every iteration by the IML estimator, so the comparison isolates the noise estimate from the step-size estimate.
For both deblurring and $2\times$ SR, the two trajectories are nearly indistinguishable and reach the same final PSNR to within $0.1$\,dB.
Thus, AdaptivePnP removes the need for known measurement noise $\sigma$ at no cost in reconstruction quality.

\section{Conclusion}

In this paper, we introduced AdaptivePnP, a framework for jointly estimating the measurement noise level $\sigma$ and PnP step size $\gamma$ directly from the data-fitting residual during reconstruction.  To estimate the parameters, we propose an IML estimator that uses the normalized Cram\'er--Rao bound to restrict estimated parameters to an identifiable window.  
As applied in AdaptivePnP, the IML estimator uses a virtual model 
in which the residual has covariance $\gamma^2 AA^{\T}+\sigma^2 I$, so $\sigma$ and $\gamma$ contribute distinct spectral components that can be jointly estimated.
Using the virtual model, we showed that this window reduces to an interval of ratios $\rho=\gamma/\sigma$, yielding a simple 2-step procedure to estimate $\sigma^2$ and $\gamma^2$.

We applied AdaptivePnP to PnP-PGM and PnP-ADMM, yielding algorithms that require neither a tuned $\gamma$ nor a known $\sigma$. 
Our experiments on image deblurring and $2\times$ SR demonstrated that AdaptivePnP matches or outperforms corresponding PnP baselines that are supplied with the true measurement noise level and use a step size selected by grid search on paired ground truth validation data.
The improvement is particularly pronounced for $2\times$ SR, where adapting the parameters during reconstruction substantially outperforms a single validation-tuned parameter choice.
Additionally, we found that the constrained IML estimator is essential for reconstruction quality.
These results suggest that explicitly accounting for parameter identifiability can provide a practical route toward PnP reconstruction methods that adapt their regularization and noise parameters directly from the observed data rather than relying on problem-specific offline tuning.

\appendix
\input{appendices}

\section*{Acknowledgment}
The authors used Claude Fable 5.1 in this work for developing and documenting code and revising the manuscript. All AI-assisted code and text were reviewed, verified, and edited by the authors, who take full responsibility for the content of this article.

\bibliographystyle{IEEEtran}
\bibliography{main}



\end{document}

%% file: appendices.tex
\section{Derivations for the IML Estimator}\label{app:derivations}

This appendix proves the results used in Section~\ref{sec:IML-application} and derives the profiled likelihood minimized in Algorithm~\ref{alg:iml}.
We write $\rho_{\inf}(\tau)$ and $\rho_{\sup}(\tau)$ for the endpoints $\rho_{\inf}$ and $\rho_{\sup}$ of Section~\ref{sec:IML-application} to make their dependence on $\tau$ explicit, and
\begin{equation}\label{eq:app-s}
  s_i = \gamma^2\lambda_i + \sigma^2 = \sigma^2\bigl(1+\rho^2\lambda_i\bigr),
  \qquad i = 1,\dots,M,
\end{equation}
for the variance of $\tilde Z_i$, so that $\sigma^2 C_\rho = \diag\{s_i\}$.

\begin{lemma}[Normalized Cram\'er--Rao Bounds]\label{lem:fisher}
The Fisher information matrix of $\theta = (\sigma^2,\gamma^2)$ under~\eqref{eq:WhitenedModel} is
\begin{equation}\label{eq:fisher-rho}
  \I(\theta) = \frac{1}{2\sigma^4}
  \begin{bmatrix}
    \Tr\bigl(C_\rho^{-2}\bigr)     & \Tr\bigl(\Lambda C_\rho^{-2}\bigr) \\
    \Tr\bigl(\Lambda C_\rho^{-2}\bigr) & \Tr\bigl(\Lambda^{2} C_\rho^{-2}\bigr)
  \end{bmatrix},
\end{equation}
and its determinant is
\begin{equation}\label{eq:app-detI}
  \det\I(\theta) = \frac{\Delta(\rho)}{4\sigma^{8}},
\end{equation}
with $\Delta(\rho)$ as in~\eqref{eq:Delta}.
In particular, $\I(\theta)$ is nonsingular if and only if $\Lambda$ is not a multiple of $I$, in which case the normalized Cram\'er--Rao bounds~\eqref{eq:v_i} of $\sigma^2$ and $\gamma^2$ are given by~\eqref{eq:Vsigma} and~\eqref{eq:Vgamma}.
\end{lemma}

\begin{proof}
Since the components of $\tilde Z$ from~\eqref{eq:WhitenedModel} are independent with $\tilde Z_i \sim \mathcal{N}(0,s_i)$, the negative log-likelihood is
\begin{equation}\label{eq:app-nll}
  f_\theta(\tilde z) = \frac12\sum_{i=1}^{M}\left(\log s_i + \frac{\tilde z_i^2}{s_i}\right),
\end{equation}
up to an additive constant. By~\eqref{eq:app-s}, $\partial s_i/\partial\sigma^2 = 1$ and $\partial s_i/\partial\gamma^2 = \lambda_i$, so the second partial derivatives of $f_\theta$ are
\begin{align}
  \frac{\partial^2 f_\theta}{\partial(\sigma^2)^2}
  &= \frac12\sum_{i=1}^{M}\left(\frac{2\tilde z_i^2}{s_i^3} - \frac{1}{s_i^2}\right),\\
  \frac{\partial^2 f_\theta}{\partial\sigma^2\,\partial\gamma^2}
  &= \frac12\sum_{i=1}^{M}\lambda_i\left(\frac{2\tilde z_i^2}{s_i^3} - \frac{1}{s_i^2}\right),\\
  \frac{\partial^2 f_\theta}{\partial(\gamma^2)^2}
  &= \frac12\sum_{i=1}^{M}\lambda_i^2\left(\frac{2\tilde z_i^2}{s_i^3} - \frac{1}{s_i^2}\right).
\end{align}
Taking expectations with $\E[\tilde Z_i^2] = s_i$ gives
\begin{equation}\label{eq:app-fim}
  \I(\theta) = \frac12
  \begin{bmatrix}
    \sum_{i} s_i^{-2} & \sum_{i} \lambda_i s_i^{-2} \\[0.5ex]
    \sum_{i} \lambda_i s_i^{-2} & \sum_{i} \lambda_i^{2} s_i^{-2}
  \end{bmatrix},
\end{equation}
and substituting $s_i^{-2} = \sigma^{-4}[C_\rho^{-2}]_{ii}$ gives~\eqref{eq:fisher-rho}.
 
From~\eqref{eq:app-fim},
\begin{equation}\label{eq:app-det}
  4\det\I(\theta)
  = \Bigl(\sum_{i} s_i^{-2}\Bigr)\Bigl(\sum_{i}\lambda_i^2 s_i^{-2}\Bigr)
    - \Bigl(\sum_{i}\lambda_i s_i^{-2}\Bigr)^{2},
\end{equation}
so again substituting $s_i^{-2} = \sigma^{-4}[C_\rho^{-2}]_{ii}$ gives~\eqref{eq:app-detI}.
By the Cauchy--Schwarz inequality applied to the vectors $(s_i^{-1})_{i}$ and $(\lambda_i s_i^{-1})_{i}$, the right-hand side of~\eqref{eq:app-det} is nonnegative, with equality if and only if the two vectors are proportional.
Since every $s_i > 0$, this happens if and only if all $\lambda_i$ are equal.
Therefore, $\I(\theta)$ is nonsingular if and only if $\Lambda$ is not a multiple of $I$.

Finally, when $\I(\theta)$ is nonsingular, inverting the $2\times 2$ matrix in~\eqref{eq:fisher-rho} gives
\begin{equation}\label{eq:fisher-inv}
  \I(\theta)^{-1} = \frac{2\sigma^4}{\Delta(\rho)}
  \begin{bmatrix}
    \Tr\bigl(\Lambda^{2} C_\rho^{-2}\bigr) & -\Tr\bigl(\Lambda C_\rho^{-2}\bigr) \\
    -\Tr\bigl(\Lambda C_\rho^{-2}\bigr)    & \Tr\bigl(C_\rho^{-2}\bigr)
  \end{bmatrix},
\end{equation}
and dividing its diagonal entries by $\gamma^4 = \rho^4\sigma^4$ and $\sigma^4$ respectively gives the normalized Cram\'er--Rao bounds $V_\gamma(\rho;\Lambda)$ and $V_\sigma(\rho;\Lambda)$ from~\eqref{eq:Vgamma} and~\eqref{eq:Vsigma}
 
\end{proof}

Now we show that $V_\gamma(\rho;\Lambda)$ is strictly decreasing and $V_\sigma(\rho;\Lambda)$ is increasing in $\rho$, so that each constraint from the identifiable window $W_\tau$ in~\eqref{eq:general-window} bounds $\rho$ on one side.
Define
\begin{align}
  \rho_{\inf}(\tau) &= \inf\{\rho > 0: V_\gamma(\rho;\Lambda) \leq \tau^2\}, \label{eq:rhoInf}\\
  \rho_{\sup}(\tau) &= \sup\{\rho > 0: V_\sigma(\rho;\Lambda) \leq \tau^2\}, \label{eq:rhoSup}
\end{align}
with the conventions $\inf\emptyset = \infty$ and $\sup\emptyset = 0$.
Then the identifiable window is the set of parameter pairs whose ratio lies between these two limits,
\begin{equation}\label{eq:model-window}
  \W_\tau = \left\{(\sigma^2,\gamma^2): \rho_{\inf}^2(\tau) \leq  \frac{\gamma^2}{\sigma^2} \leq \rho_{\sup}^2(\tau) \right\},
\end{equation}
which is the interval~\eqref{eq:rho-interval} of Section~\ref{sec:IML-application}, and is empty if $\rho_{\inf}(\tau) = \infty$ or $\rho_{\inf}(\tau) > \rho_{\sup}(\tau)$.

\begin{theorem}[Monotonicity and the Identifiable Window]\label{thm:mono}
Assume that $\Lambda$ is not a multiple of $I$, and let $M_0$ be the number of zero eigenvalues of $AA^\T$.
Then $V_\sigma(\rho;\Lambda)$ and $V_\gamma(\rho;\Lambda)$ are continuous on $(0,\infty)$, and the following hold.
\begin{enumerate}
  \item[(i)] $V_\gamma(\rho;\Lambda)$ is strictly decreasing on $(0,\infty)$, $V_\gamma(\rho;\Lambda) \to \infty$ as $\rho \to 0$, and
  \begin{equation}\label{eq:VgammaInf}
    \lim_{\rho\to\infty} V_\gamma(\rho;\Lambda) =
    \begin{cases}
      \dfrac{2}{M - M_0}, & M_0 \geq 1,\\[1.5ex]
      \frac{2\Tr(\Lambda^{-2})}{M\Tr(\Lambda^{-2}) - [\Tr\Lambda^{-1}]^2}, & M_0 = 0.
    \end{cases}
  \end{equation}
  \item[(ii)] $V_\sigma(\rho;\Lambda)$ is increasing on $(0,\infty)$, strictly increasing if $\Lambda$ has at least two distinct nonzero eigenvalues, and
  \begin{align}
    \lim_{\rho\to 0} V_\sigma(\rho;\Lambda) &= \frac{2\Tr(\Lambda^2)}{M\Tr(\Lambda^2) - [\Tr\Lambda]^2}, \label{eq:Vsigma0}\\
    \lim_{\rho\to\infty} V_\sigma(\rho;\Lambda) &=
    \begin{cases}
      2/M_0, & M_0 \geq 1,\\[0.5ex]
      \infty, & M_0 = 0.
    \end{cases} \label{eq:VsigmaInf}
  \end{align}
\end{enumerate}
Consequently, for every $\tau > 0$, $\W_\tau$ is given by~\eqref{eq:model-window}, and it is empty if and only if $\rho_{\inf}(\tau) = \infty$ or $\rho_{\inf}(\tau) > \rho_{\sup}(\tau)$.
\end{theorem}

\begin{proof}
For $\rho > 0$, define
\begin{equation}\label{eq:noise-fraction}
  x_i = \frac{1}{1+\rho^2\lambda_i}, \quad
  y_i = 1 - x_i.
\end{equation}
Additionally, let 
\begin{equation}
    T_1 = \sum_i x_i, \quad T_2 = \sum_i x_i^2
\end{equation}
and 
\begin{equation}
    S_1 = \sum_i y_i, \quad S_2 = \sum_i y_i^2.
\end{equation}
Observe that $C_\rho^{-1} = \diag\{x_i\}$ and $y_i = \rho^2\lambda_i x_i$.
Thus,
\begin{equation}
\begin{aligned}
  \Tr(C_\rho^{-2}) &= T_2, \\
  \Tr(\Lambda C_\rho^{-2}) &= \rho^{-2}(T_1 - T_2),\\
  \Tr(\Lambda^2 C_\rho^{-2}) &= \rho^{-4} S_2.
\end{aligned}
\end{equation}
Substituting into $\Delta(\rho)$ given in~\eqref{eq:Delta} and using $S_2 = M - 2T_1 + T_2$ and $S_1 = M - T_1$,
\begin{equation}\label{eq:rho4Delta}
\begin{split}
  \rho^4\Delta(\rho) &= T_2(M - 2T_1 + T_2) - (T_1 - T_2)^2\\
  &= M T_2 - T_1^2 = M S_2 - S_1^2 > 0,
\end{split}
\end{equation}
where positivity follows from Lemma~\ref{lem:fisher}.
By definition of $V_\gamma(\rho;\Lambda)$ and $V_\sigma(\rho;\Lambda)$ given in~\eqref{eq:Vgamma} and~\eqref{eq:Vsigma},
\begin{align}
  V_\gamma(\rho;\Lambda) &= \frac{2}{M - T_1^2/T_2},\label{eq:V_gam}\\
  \qquad
   V_\sigma(\rho;\Lambda) &= \frac{2}{M - S_1^2/S_2}.\label{eq:V_sig}
\end{align}
Each $x_i$ is continuous in $\rho$, and the denominators in~\eqref{eq:V_gam} and~\eqref{eq:V_sig} are positive.
Thus, $V_\sigma(\rho;\Lambda)$ and $V_\gamma(\rho;\Lambda)$ are continuous on $(0,\infty)$.

First we will prove (i).
By~\eqref{eq:V_gam}, $V_\gamma(\rho;\Lambda)$ is strictly decreasing if and only if $T_1^2/T_2$ is strictly decreasing.
Differentiating with respect to $\rho$, we have
\begin{equation}\label{eq:dT}
  \frac{d}{d\rho}\frac{T_1^2}{T_2}
  = \frac{4\rho T_1}{T_2^2}\sum_{i,j}\lambda_i x_i^2 x_j (x_i - x_j).
\end{equation}
Relabeling the summation indices $i \leftrightarrow j$ leaves the sum in~\eqref{eq:dT} unchanged, so it equals the average of itself and its relabeled copy, i.e.,
\begin{equation}\label{eq:sym}
\begin{split}
  \sum_{i,j}\lambda_i x_i^2 x_j (x_i - x_j)
  &= \frac12\sum_{i,j}\bigl(\lambda_i x_i^2 x_j - \lambda_j x_j^2 x_i\bigr)(x_i - x_j)\\
  &= \frac12\sum_{i,j} x_i x_j\,(x_i - x_j)(\lambda_i x_i - \lambda_j x_j).
\end{split}
\end{equation}
Using the definition of $x_i$, for every pair $(i,j)$, we have
\begin{equation}\label{eq:pair-ids}
  -\rho^{-2}(x_i - x_j) = \lambda_i x_i - \lambda_j x_j = (\lambda_i - \lambda_j)\,x_i x_j.
\end{equation}
Substituting~\eqref{eq:sym} and then~\eqref{eq:pair-ids} into~\eqref{eq:dT},
\begin{equation}
\begin{split}
  \frac{d}{d\rho}\frac{T_1^2}{T_2}
  &= \frac{2\rho T_1}{T_2^2}\sum_{i,j} x_i x_j (x_i - x_j)(\lambda_i x_i - \lambda_j x_j)\\
  &= -\frac{2\rho^3 T_1}{T_2^2}\sum_{i,j} (\lambda_i - \lambda_j)^2\, x_i^3 x_j^3 < 0,
\end{split}
\end{equation}
where negativity follows from $x_i > 0$ and $\lambda_i \neq \lambda_j$ for at least one pair.
Additionally, $\lim_{\rho \to 0} x_i = 1$ for all $i$, so $T_1^2/T_2 \to M$.
Hence,
\begin{equation}
    \lim_{\rho\to 0}V_\gamma(\rho;\Lambda) = \infty.
\end{equation}
Now consider $\rho \to \infty$.
If $M_0 \geq 1$, then
\begin{equation}
\lim_{\rho\to \infty}x_i =
    \begin{cases}
        1 &\text{if } \lambda_i = 0 \\
        0 &\text{else.}
    \end{cases}
\end{equation}
Thus, $T_1^2/T_2 \to M_0$ and
\begin{equation}
    \lim_{\rho\to \infty}V_\gamma(\rho;\Lambda) = \frac{2}{M-M_0}.
\end{equation}
On the other hand, if there are no zero eigenvalues ($M_0 = 0$), then $\rho^2 x_i \to 1/\lambda_i$ for every $i$. Thus, $T_1^2/T_2 \to (\sum_i \lambda_i^{-1})^2/\sum_i \lambda_i^{-2}$ and 
\begin{equation}
    \lim_{\rho\to \infty}V_\gamma(\rho;\Lambda) = \frac{2\Tr(\Lambda^{-2})}{(M\Tr(\Lambda^{-2}) - [\Tr\Lambda^{-1}]^2)}.
\end{equation}
This proves~\eqref{eq:VgammaInf}.

Next, we will prove (ii).
By~\eqref{eq:V_sig}, $V_\sigma(\rho;\Lambda)$ is increasing if and only if $S_1^2/S_2$ is increasing.
Differentiating with respect to $\rho$ and substituting $y_j - y_i = x_i - x_j$ and $y_j = \rho^2\lambda_j x_j$, we have
\begin{equation}
\begin{split}
  \frac{d}{d\rho}\frac{S_1^2}{S_2}
  &= \frac{4\rho S_1}{S_2^2}\sum_{i,j}\lambda_i x_i^2\, y_j\,(y_j - y_i)\\
  &= \frac{4\rho^3 S_1}{S_2^2}\sum_{i,j}\lambda_i\lambda_j\, x_i^2 x_j\,(x_i - x_j).
\end{split}
\end{equation}
Symmetrizing as in~\eqref{eq:sym} and applying~\eqref{eq:pair-ids}, we have
\begin{equation}
\begin{split}
  \frac{d}{d\rho}\frac{S_1^2}{S_2}
  &= \frac{2\rho^3 S_1}{S_2^2}\sum_{i,j}\lambda_i\lambda_j\, x_i x_j\,(x_i - x_j)^2\\
  &= \frac{2\rho^7 S_1}{S_2^2}\sum_{i,j}\lambda_i\lambda_j(\lambda_i - \lambda_j)^2\, x_i^3 x_j^3 \geq 0.
\end{split}
\end{equation}
Each term in this sum is positive exactly when $\lambda_i,\lambda_j > 0$ and $\lambda_i \neq \lambda_j$.
Thus, $V_\sigma(\rho;\Lambda)$ is increasing, and is strictly increasing if and only if $\Lambda$ has at least two distinct nonzero eigenvalues.
Additionally, as $\rho \to 0$, $y_i/\rho^2 \to \lambda_i$ for every $i$, so $S_1^2/S_2 \to (\sum_i \lambda_i)^2/\sum_i \lambda_i^2$. Hence,
\begin{equation}
    \lim_{\rho \to 0} V_\sigma(\rho;\Lambda) = \frac{2\Tr(\Lambda^2)}{(M\Tr(\Lambda^2) - [\Tr\Lambda]^2)}.
\end{equation}
This proves~\eqref{eq:Vsigma0}.
Now consider $\rho \to \infty$. We know
\begin{equation}
    \lim_{\rho\to \infty}y_i =
    \begin{cases}
        0 &\text{if } \lambda_i = 0 \\
        1 &\text{else.}
    \end{cases}
\end{equation}
Thus, $S_1^2/S_2 \to M - M_0$ and
\begin{equation}
    \lim_{\rho\to \infty} V_\sigma(\rho;\Lambda) = \begin{cases}
        2/M_0 &\text{if \ } M_0 > 0 \\
        \infty &\text{if \ } M_0 = 0.
    \end{cases}
\end{equation}
This proves~\eqref{eq:VsigmaInf}.

Finally, fix $\tau > 0$ and let
\begin{align}
  L_\gamma &= \{\rho > 0 : V_\gamma(\rho;\Lambda) \leq \tau^2\},\label{eq:level-set-gamma}\\
  L_\sigma &= \{\rho > 0 : V_\sigma(\rho;\Lambda) \leq \tau^2\} \label{eq:level-set-sigma}.
\end{align}
Then, as defined in~\eqref{eq:rhoInf} and~\eqref{eq:rhoSup}, $\rho_{\inf}(\tau) = \inf L_\gamma$ and $\rho_{\sup}(\tau) = \sup L_\sigma$, and $\W_\tau$ is the set of $(\sigma^2,\gamma^2)$ with $\gamma/\sigma \in L_\gamma \cap L_\sigma$.
Both sets are closed in $(0,\infty)$ because $V_\gamma(\rho;\Lambda)$ and $V_\sigma(\rho;\Lambda)$ are continuous.
Using monotonicity and the limits in (i) and (ii), we have
\begin{equation}\label{eq:set_gamma}
  L_\gamma =
  \begin{cases}
    \emptyset, &\text{if\ } \rho_{\inf}(\tau) = \infty,\\
    [\rho_{\inf}(\tau), \infty), &\text{if\ } \rho_{\inf}(\tau) < \infty,
  \end{cases}
\end{equation}
 and 
\begin{equation}\label{eq:set_sigma}
   L_\sigma =
  \begin{cases}
    \emptyset, &\text{if\ } \rho_{\sup}(\tau) = 0,\\
    (0, \rho_{\sup}(\tau)], &\text{if\ } 0 < \rho_{\sup}(\tau) < \infty,\\
    (0,\infty), &\text{if\ } \rho_{\sup}(\tau) = \infty.
  \end{cases}
\end{equation}
Therefore, $L_\gamma \cap L_\sigma = \{\rho > 0 : \rho_{\inf}(\tau) \leq \rho \leq \rho_{\sup}(\tau)\}$ as in~\eqref{eq:model-window}, and this set is empty if and only if $\rho_{\inf}(\tau) = \infty$ or $\rho_{\inf}(\tau) > \rho_{\sup}(\tau)$.
\end{proof}

Theorem~\ref{thm:mono} shows that $\W_\tau$ is empty when $\tau$ is too small; the following corollary gives the exact threshold for $\tau$ such that $\W_\tau$ is nonempty.
\begin{corollary}[Smallest Tolerance for Nonempty Window]\label{cor:tau-star}
Assume that $\Lambda$ is not a multiple of $I$ and that fewer than half of its eigenvalues are zero.
Then $V_\gamma(\rho;\Lambda)$ and $V_\sigma(\rho;\Lambda)$ cross at a unique $\rho_\star > 0$, and $\W_\tau$ is nonempty if and only if
\begin{equation}\label{eq:tau-star}
  \tau^2 \geq \tau_\star^2 = V_\gamma(\rho_\star;\Lambda) = V_\sigma(\rho_\star;\Lambda).
\end{equation}
\end{corollary}

\begin{proof}
By Theorem~\ref{thm:mono}, $V_\gamma(\rho;\Lambda) - V_\sigma(\rho;\Lambda)$ is continuous and strictly decreasing on $(0,\infty)$ and tends to $\infty$ as $\rho \to 0$.
Additionally, by~\eqref{eq:VgammaInf} and~\eqref{eq:VsigmaInf}, its limit as $\rho \to \infty$ is negative when $M_0 < M/2$.
Thus, there exists a unique $\rho_\star > 0$ such that $(V_\gamma - V_\sigma)(\rho_\star;\Lambda) = 0$.

Since $V_\gamma(\rho;\Lambda)$ is decreasing and $V_\sigma(\rho;\Lambda)$ is increasing, we know that 
\begin{equation}
    \max\{V_\gamma(\rho;\Lambda), V_\sigma(\rho;\Lambda)\} \geq \tau_\star^2
\end{equation}
for every $\rho$, with equality at $\rho_\star$.
By~\eqref{eq:general-window}, $\W_\tau$ is nonempty if and only if some $\rho$ satisfies $\max\{V_\gamma(\rho;\Lambda), V_\sigma(\rho;\Lambda)\} \leq \tau^2$, which holds if and only if $\tau^2 \geq \tau_\star^2$.
\end{proof}

The negative log-likelihood of $\tilde{z}$ under~\eqref{eq:WhitenedModel} is
\begin{equation}
  -\log p_\theta(\tilde z) = \frac{M}{2}\log \sigma^2 + \frac12 \log\det C_\rho + \frac{1}{2\sigma^2}\, \tilde{z}^\T C_\rho^{-1} \tilde{z},
  \label{eq:NegLogLik}
\end{equation}
up to an additive constant independent of $\theta$.
Setting the derivative with respect to $\sigma^2$ to zero gives the ML estimate of $\sigma^2$ for fixed $\rho$, i.e.,
\begin{equation}
  \hat{\sigma}^2(\rho) = \frac{1}{M}\, \tilde{z}^\T C_\rho^{-1} \tilde{z}.
  \label{eq:SigmaHatRho}
\end{equation}
Note that $\hat{\sigma}^2(\rho)$ is positive whenever $\tilde{z} \neq 0$.
Substituting~\eqref{eq:SigmaHatRho} into~\eqref{eq:NegLogLik} gives $\tfrac12\mathcal{L}(\rho;\tilde z,\Lambda)$ plus an additive constant, with $\mathcal{L}$ as in~\eqref{eq:Profile}.
Since $\W_\tau$ constrains only $\rho$, the IML estimate of $\rho$ is
\begin{equation}
  \hat{\rho} = \argmin_{\rho \in [\rho_{\inf}(\tau), \rho_{\sup}(\tau)]} \mathcal{L}(\rho; \tilde{z}),
  \label{eq:RhoHat}
\end{equation}
and substituting $\hat{\rho}$ into~\eqref{eq:SigmaHatRho} gives the last two steps of Algorithm~\ref{alg:iml}.

\begin{theorem}[Existence of the IML Estimate]\label{thm:exist}
Assume that $\W_\tau$ is nonempty, $\tilde z \neq 0$, and $\Lambda$ is not a multiple of $I$.
If $\rho_{\sup}(\tau) = \infty$, assume additionally that $\tilde z_i \neq 0$ for some $i$ with $\lambda_i = 0$.
Then, the minimum in~\eqref{eq:RhoHat} is attained.
\end{theorem}

\begin{proof}
Let 
\begin{equation}
    J = \{\rho > 0 : \rho_{\inf}(\tau) \leq \rho \leq \rho_{\sup}(\tau)\}.
\end{equation}
By the identifiable window $W_\tau$ given in~\eqref{eq:model-window}, $(\sigma^2,\gamma^2) \in \W_\tau$ if and only if $\gamma/\sigma \in J$.
We want to show that $\mathcal{L}(\rho;\tilde z, \Lambda)$ in~\eqref{eq:Profile} attains its minimum on $J$.
Since $\tilde z \neq 0$ and $C_\rho^{-1}$ is positive definite, $\tilde z^\T C_\rho^{-1}\tilde z > 0$ for every $\rho > 0$, so $\mathcal{L}(\rho;\tilde z, \Lambda)$ is continuous on $(0,\infty)$.
Additionally, since $\W_\tau \neq \emptyset$, Theorem~\ref{thm:mono} gives $\rho_{\inf}(\tau) < \infty$ and $\rho_{\inf}(\tau) \leq \rho_{\sup}(\tau)$, so $J \neq \emptyset$.
Moreover, $\rho_{\inf}(\tau) > 0$ since $\lim_{\rho \to 0}V_\gamma = \infty$.
Thus, if $\rho_{\sup}(\tau) < \infty$, $J$ is compact and, therefore, the minimum in~\eqref{eq:RhoHat} is attained.

On the other hand, if $\rho_{\sup}(\tau) = \infty$, then $J$ is the set of $\rho$ in $[\rho_{\inf}(\tau), \infty)$. 
Since $\lim_{\rho \to \infty} V_\sigma(\rho;\Lambda) \leq \tau^2 < \infty$, we know $M_0 \geq 1$ by~\eqref{eq:VsigmaInf}. Additionally, $M_0 < M$ since $\Lambda$ is not a multiple of $I$.
Since $[C_\rho^{-1}]_{ii} = (1 + \rho^2\lambda_i)^{-1}$ tends to $0$ if $\lambda_i \\
> 0$ and equals $1$ if $\lambda_i = 0$, we know
\begin{equation}
  \lim_{\rho \to \infty} \tilde z^\T C_\rho^{-1}\tilde z = \sum_{i:\,\lambda_i = 0}\tilde z_i^2 > 0,
\end{equation}
where positivity holds since $\tilde{z} \neq 0$.
Additionally, since we have assumed at least one $\lambda_i$ is positive,
\begin{equation}
  \lim_{\rho \to \infty} \log\det C_\rho = \lim_{\rho \to \infty}  \sum_{i:\,\lambda_i > 0}\log\bigl(1 + \rho^2\lambda_i\bigr) = \infty.
\end{equation}
Thus,
\begin{equation}
    \lim_{\rho \to \infty} \mathcal{L}(\rho;\tilde z, \Lambda) = \infty,
\end{equation} so the sublevel set $\{\rho \in J : \mathcal{L}(\rho;\tilde z, \Lambda) \leq \mathcal{L}(\rho_{\inf}(\tau);\tilde z)\}$ is nonempty, closed, and bounded, hence compact.
Therefore, the minimum in~\eqref{eq:RhoHat} is attained.
\end{proof}

%% file: main.bib
@inproceedings{venkatakrishnan_2013_plugandplay,
    title = {Plug-and-{Play} priors for model based reconstruction},
    doi = {10.1109/GlobalSIP.2013.6737048},
    booktitle = {2013 {IEEE} {Global} {Conference} on {Signal} and {Information} {Processing}},
    author = {Venkatakrishnan, Singanallur V. and Bouman, Charles A. and Wohlberg, Brendt},
    month = dec,
    year = {2013},
    pages = {945--948},
}

@article{zhang_2021_dpir,
  title={Plug-and-play image restoration with deep denoiser prior},
  author={Zhang, Kai and Li, Yawei and Zuo, Wangmeng and Zhang, Lei and Van Gool, Luc and Timofte, Radu},
  journal={IEEE Transactions on Pattern Analysis and Machine Intelligence},
  volume={44},
  number={10},
  pages={6360--6376},
  year={2021},
  publisher={IEEE}
}

@book{brent2013algorithms,
  title={Algorithms for minimization without derivatives},
  author={Brent, Richard P},
  year={2013},
  publisher={Courier Corporation}
}

@article{hurault_2022_gspnp,
  title={Gradient step denoiser for convergent plug-and-play},
  author={Hurault, Samuel and Leclaire, Arthur and Papadakis, Nicolas},
  journal={arXiv preprint arXiv:2110.03220},
  year={2021}
}

@article{kamilov_2023_plugandplay,
    title = {Plug-and-{Play} {Methods} for {Integrating} {Physical} and {Learned} {Models} in {Computational} {Imaging}: {Theory}, algorithms, and applications},
    volume = {40},
    issn = {1558-0792},
    shorttitle = {Plug-and-{Play} {Methods} for {Integrating} {Physical} and {Learned} {Models} in {Computational} {Imaging}},
    doi = {10.1109/MSP.2022.3199595},
    number = {1},
    journal = {IEEE Signal Processing Magazine},
    author = {Kamilov, Ulugbek S. and Bouman, Charles A. and Buzzard, Gregery T. and Wohlberg, Brendt},
    month = jan,
    year = {2023},
    pages = {85--97},
}

@article{wohlberg_2017_admm,
  title={ADMM penalty parameter selection by residual balancing},
  author={Wohlberg, Brendt},
  journal={arXiv preprint arXiv:1704.06209},
  year={2017}
}

@inproceedings{xu_2017_relaxedadmm,
  title={Adaptive relaxed ADMM: Convergence theory and practical implementation},
  author={Xu, Zheng and Figueiredo, Mario AT and Yuan, Xiaoming and Studer, Christoph and Goldstein, Tom},
  booktitle={2017 IEEE Conference on Computer Vision and Pattern Recognition (CVPR)},
  pages={7234--7243},
  year={2017},
  organization={IEEE}
}

@inproceedings{zhang_2018_istaNet,
    title = {{ISTA}-{Net}: {Interpretable} {Optimization}-{Inspired} {Deep} {Network} for {Image} {Compressive} {Sensing}},
    issn = {2575-7075},
    shorttitle = {{ISTA}-{Net}},
    doi = {10.1109/CVPR.2018.00196},
    booktitle = {2018 {IEEE}/{CVF} {Conference} on {Computer} {Vision} and {Pattern} {Recognition}},
    author = {Zhang, Jian and Ghanem, Bernard},
    month = jun,
    year = {2018},
    note = {ISSN: 2575-7075},
    pages = {1828--1837},
}

@inproceedings{yang_2016_admmNet,
    title = {Deep {ADMM}-{Net} for {Compressive} {Sensing} {MRI}},
    volume = {29},
    booktitle = {Advances in {Neural} {Information} {Processing} {Systems}},
    publisher = {Curran Associates, Inc.},
    author = {yang, yan and Sun, Jian and Li, Huibin and Xu, Zongben},
    year = {2016},
}

@article{wei_2022_tfpnp,
    title = {{TFPnP}: {Tuning}-free {Plug}-and-{Play} {Proximal} {Algorithms} with {Applications} to {Inverse} {Imaging} {Problems}},
    volume = {23},
    issn = {1533-7928},
    shorttitle = {{TFPnP}},
    number = {16},
    journal = {Journal of Machine Learning Research},
    author = {Wei, Kaixuan and Aviles-Rivero, Angelica and Liang, Jingwei and Fu, Ying and Huang, Hua and Schönlieb, Carola-Bibiane},
    year = {2022},
    pages = {1--48},
}

@article{li_2025_curvpnp,
    title = {{CurvPnP}: {Plug}-and-play blind image restoration with deep curvature denoiser},
    volume = {233},
    issn = {0165-1684},
    shorttitle = {{CurvPnP}},
    doi = {10.1016/j.sigpro.2025.109951},
    journal = {Signal Processing},
    author = {Li, Yutong and Chang, Huibin and Duan, Yuping},
    month = aug,
    year = {2025},
    pages = {109951},
}

@article{liu_2025_plugandplay,
    title = {Plug-and-{Play} {ADMM} for {Embedded} {Noise} {Level} {Estimation}},
    volume = {67},
    issn = {0924-9907, 1573-7683},
    doi = {10.1007/s10851-025-01256-8},
    language = {en},
    number = {4},
    journal = {Journal of Mathematical Imaging and Vision},
    author = {Liu, Hanxin and Fang, Zhuang and Tang, Liming and Lu, Wenjing},
    month = aug,
    year = {2025},
    pages = {36},
}

@article{morozov_1968_discrepancy,
    title = {The error principle in the solution of operational equations by the regularization method},
    volume = {8},
    issn = {0041-5553},
    doi = {10.1016/0041-5553(68)90034-7},
    number = {2},
    journal = {USSR Computational Mathematics and Mathematical Physics},
    publisher = {Elsevier BV},
    author = {Morozov, V.A.},
    month = jan,
    year = {1968},
    pages = {63--87},
}

@article{golub_1979_gcv,
    title = {Generalized {Cross}-{Validation} as a {Method} for {Choosing} a {Good} {Ridge} {Parameter}},
    volume = {21},
    issn = {0040-1706},
    doi = {10.1080/00401706.1979.10489751},
    number = {2},
    journal = {Technometrics},
    publisher = {Informa UK Limited},
    author = {Golub, Gene H. and Heath, Michael and Wahba, Grace},
    month = may,
    year = {1979},
    pages = {215--223},
}

@article{stein_1981_sure,
    title = {Estimation of the {Mean} of a {Multivariate} {Normal} {Distribution}},
    volume = {9},
    issn = {0090-5364},
    doi = {10.1214/aos/1176345632},
    number = {6},
    journal = {The Annals of Statistics},
    publisher = {Institute of Mathematical Statistics},
    author = {Stein, Charles M.},
    month = nov,
    year = {1981},
}

@article{hansen_1993_use,
    title = {The {Use} of the {L}-{Curve} in the {Regularization} of {Discrete} {Ill}-{Posed} {Problems}},
    volume = {14},
    issn = {1064-8275},
    doi = {10.1137/0914086},
    number = {6},
    journal = {SIAM Journal on Scientific Computing},
    publisher = {Society for Industrial and Applied Mathematics},
    author = {Hansen, Per Christian and O’Leary, Dianne Prost},
    month = nov,
    year = {1993},
    pages = {1487--1503},
}

@article{he_2000,
    title = {Alternating {Direction} {Method} with {Self}-{Adaptive} {Penalty} {Parameters} for {Monotone} {Variational} {Inequalities}},
    volume = {106},
    copyright = {https://www.springernature.com/gp/researchers/text-and-data-mining},
    issn = {0022-3239, 1573-2878},
    doi = {10.1023/A:1004603514434},
    language = {en},
    number = {2},
    journal = {Journal of Optimization Theory and Applications},
    author = {He, B. S. and Yang, H. and Wang, S. L.},
    month = aug,
    year = {2000},
    pages = {337--356},
}

@inproceedings{xu_2017_adaptivepenalty,
    title = {{ADMM} without a {Fixed} {Penalty} {Parameter}: {Faster} {Convergence} with {New} {Adaptive} {Penalization}},
    volume = {30},
    shorttitle = {{ADMM} without a {Fixed} {Penalty} {Parameter}},
    booktitle = {Advances in {Neural} {Information} {Processing} {Systems}},
    publisher = {Curran Associates, Inc.},
    author = {Xu, Yi and Liu, Mingrui and Lin, Qihang and Yang, Tianbao},
    year = {2017},
}

@article{barzilai_1988,
    title = {Two-{Point} {Step} {Size} {Gradient} {Methods}},
    volume = {8},
    doi = {10.1093/imanum/8.1.141},
    number = {1},
    journal = {IMA Journal of Numerical Analysis},
    author = {Barzilai, Jonathan and Borwein, Jonathan M.},
    year = {1988},
    pages = {141--148},
}

@inproceedings{malitsky_2024_adaptivepgm,
    title = {Adaptive {Proximal} {Gradient} {Method} for {Convex} {Optimization}},
    volume = {37},
    doi = {10.52202/079017-3193},
    booktitle = {Advances in {Neural} {Information} {Processing} {Systems}},
    publisher = {Curran Associates, Inc.},
    author = {Malitsky, Yura and Mishchenko, Konstantin},
    year = {2024},
    pages = {100670--100697},
}

@inproceedings{kamilov_2012_approximate,
    title = {Approximate {Message} {Passing} with {Consistent} {Parameter} {Estimation} and {Applications} to {Sparse} {Learning}},
    volume = {25},
    booktitle = {Advances in {Neural} {Information} {Processing} {Systems}},
    publisher = {Curran Associates, Inc.},
    author = {Kamilov, Ulugbek and Rangan, Sundeep and Unser, Michael and Fletcher, Alyson K},
    year = {2012},
}

@inproceedings{fletcher_2017_learning,
    title = {Learning and free energies for vector approximate message passing},
    issn = {2379-190X},
    doi = {10.1109/ICASSP.2017.7952957},
    booktitle = {2017 {IEEE} {International} {Conference} on {Acoustics}, {Speech} and {Signal} {Processing} ({ICASSP})},
    author = {Fletcher, Alyson K. and Schniter, Philip},
    month = mar,
    year = {2017},
    note = {ISSN: 2379-190X},
    pages = {4247--4251},
}

@article{metzler_2016_denoising,
    title = {From {Denoising} to {Compressed} {Sensing}},
    volume = {62},
    issn = {1557-9654},
    doi = {10.1109/TIT.2016.2556683},
    number = {9},
    journal = {IEEE Transactions on Information Theory},
    author = {Metzler, Christopher A. and Maleki, Arian and Baraniuk, Richard G.},
    month = sep,
    year = {2016},
    pages = {5117--5144},
}

@inproceedings{karuppasamy_2024_distributed,
    title = {Distributed {Vector} {Approximate} {Message} {Passing}},
    issn = {2379-190X},
    doi = {10.1109/ICASSP48485.2024.10446383},
    booktitle = {{ICASSP} 2024 - 2024 {IEEE} {International} {Conference} on {Acoustics}, {Speech} and {Signal} {Processing} ({ICASSP})},
    author = {Karuppasamy, Mukilan and Akrout, Mohamed and Bellili, Faouzi and Mezghani, Amine},
    month = apr,
    year = {2024},
    note = {ISSN: 2379-190X},
    pages = {9611--9615},
}

@article{almeida_2013_parameter,
    title = {Parameter {Estimation} for {Blind} and {Non}-{Blind} {Deblurring} {Using} {Residual} {Whiteness} {Measures}},
    volume = {22},
    issn = {1941-0042},
    doi = {10.1109/TIP.2013.2257810},
    number = {7},
    journal = {IEEE Transactions on Image Processing},
    author = {Almeida, Mariana S. C. and Figueiredo, Mário A. T.},
    month = jul,
    year = {2013},
    pages = {2751--2763},
}

@article{lanza_2020_whiteness,
    title = {Residual {Whiteness} {Principle} for {Parameter}-{Free} {Image} {Restoration}},
    volume = {53},
    doi = {10.1553/etna_vol53s329},
    journal = {Electronic Transactions on Numerical Analysis},
    author = {Lanza, Alessandro and Pragliola, Monica and Sgallari, Fiorella},
    year = {2020},
    pages = {329--351},
}

@article{pragliola_2023_whiteness,
    title = {{ADMM}-{Based} {Residual} {Whiteness} {Principle} for {Automatic} {Parameter} {Selection} in {Single} {Image} {Super}-{Resolution} {Problems}},
    volume = {65},
    issn = {1573-7683},
    doi = {10.1007/s10851-022-01110-1},
    language = {en},
    number = {1},
    journal = {Journal of Mathematical Imaging and Vision},
    author = {Pragliola, Monica and Calatroni, Luca and Lanza, Alessandro and Sgallari, Fiorella},
    month = jan,
    year = {2023},
    pages = {99--123},
}

@article{wang_2023_hyperspectral,
    title = {Tuning-{Free} {Plug}-and-{Play} {Hyperspectral} {Image} {Deconvolution} {With} {Deep} {Priors}},
    volume = {61},
    doi = {10.1109/TGRS.2023.3253549},
    journal = {IEEE Transactions on Geoscience and Remote Sensing},
    author = {Wang, Xiuheng and Chen, Jie and Richard, Cédric},
    year = {2023},
    pages = {1--13},
}

@inproceedings{vila_2015_adaptive,
    title = {Adaptive damping and mean removal for the generalized approximate message passing algorithm},
    issn = {2379-190X},
    doi = {10.1109/ICASSP.2015.7178325},
    booktitle = {2015 {IEEE} {International} {Conference} on {Acoustics}, {Speech} and {Signal} {Processing} ({ICASSP})},
    author = {Vila, Jeremy and Schniter, Philip and Rangan, Sundeep and Krzakala, Florent and Zdeborová, Lenka},
    month = apr,
    year = {2015},
    note = {ISSN: 2379-190X},
    pages = {2021--2025},
}

@article{rangan_2019_convergence,
    title = {On the {Convergence} of {Approximate} {Message} {Passing} {With} {Arbitrary} {Matrices}},
    volume = {65},
    issn = {1557-9654},
    doi = {10.1109/TIT.2019.2913109},
    number = {9},
    journal = {IEEE Transactions on Information Theory},
    author = {Rangan, Sundeep and Schniter, Philip and Fletcher, Alyson K. and Sarkar, Subrata},
    month = sep,
    year = {2019},
    pages = {5339--5351},
}

@article{galatsanos_1992_methods,
    title = {Methods for choosing the regularization parameter and estimating the noise variance in image restoration and their relation},
    volume = {1},
    issn = {1941-0042},
    doi = {10.1109/83.148606},
    number = {3},
    journal = {IEEE Transactions on Image Processing},
    author = {Galatsanos, N.P. and Katsaggelos, A.K.},
    month = jul,
    year = {1992},
    pages = {322--336},
}

@article{mackay_1992_bayesian,
    title = {Bayesian {Interpolation}},
    volume = {4},
    issn = {0899-7667},
    doi = {10.1162/neco.1992.4.3.415},
    number = {3},
    journal = {Neural Computation},
    author = {MacKay, David J. C.},
    month = may,
    year = {1992},
    pages = {415--447},
}

@article{molina_1999_bayesian,
    title = {Bayesian and regularization methods for hyperparameter estimation in image restoration},
    volume = {8},
    issn = {1941-0042},
    doi = {10.1109/83.743857},
    number = {2},
    journal = {IEEE Transactions on Image Processing},
    author = {Molina, R. and Katsaggelos, A.K. and Mateos, J.},
    month = feb,
    year = {1999},
    pages = {231--246},
}

@article{vidal_2020_maximum,
    title = {Maximum {Likelihood} {Estimation} of {Regularization} {Parameters} in {High}-{Dimensional} {Inverse} {Problems}: {An} {Empirical} {Bayesian} {Approach} {Part} {I}: {Methodology} and {Experiments}},
    volume = {13},
    shorttitle = {Maximum {Likelihood} {Estimation} of {Regularization} {Parameters} in {High}-{Dimensional} {Inverse} {Problems}},
    doi = {10.1137/20M1339829},
    number = {4},
    journal = {SIAM Journal on Imaging Sciences},
    publisher = {Society for Industrial and Applied Mathematics},
    author = {Vidal, Ana Fernandez and De Bortoli, Valentin and Pereyra, Marcelo and Durmus, Alain},
    month = jan,
    year = {2020},
    pages = {1945--1989},
}

@article{mbakam_2024_marginal,
    title = {Marginal {Likelihood} {Estimation} in {Semiblind} {Image} {Deconvolution}: {A} {Stochastic} {Approximation} {Approach}},
    volume = {17},
    shorttitle = {Marginal {Likelihood} {Estimation} in {Semiblind} {Image} {Deconvolution}},
    doi = {10.1137/23M1584496},
    number = {2},
    journal = {SIAM Journal on Imaging Sciences},
    publisher = {Society for Industrial and Applied Mathematics},
    author = {Mbakam, Charlesquin Kemajou and Pereyra, Marcelo and Giovannelli, Jean-François},
    month = jun,
    year = {2024},
    pages = {1206--1254},
}

@article{dubasullivan_2026_PPM,
  title={Deep Pseudo-Proximal Map: A Self-Supervised Data-Fitting Agent for Iterative Reconstruction},
  author={Duba-Sullivan, Haley and Reid, Emma J and Bouman, Charles A and Buzzard, Gregery T},
  journal={arXiv preprint arXiv:2609.30727},
  year={2026}
}

@article{zhao_2016_fast,
    title = {Fast {Single} {Image} {Super}-{Resolution} {Using} a {New} {Analytical} {Solution} for $\ell_2$--$\ell_2$ {Problems}},
    volume = {25},
    issn = {1941-0042},
    doi = {10.1109/TIP.2016.2567075},
    number = {8},
    journal = {IEEE Transactions on Image Processing},
    author = {Zhao, Ningning and Wei, Qi and Basarab, Adrian and Dobigeon, Nicolas and Kouamé, Denis and Tourneret, Jean-Yves},
    month = aug,
    year = {2016},
    pages = {3683--3697},
}

@article{jacquez_1985_numerical,
    title = {Numerical parameter identifiability and estimability: {Integrating} identifiability, estimability, and optimal sampling design},
    volume = {77},
    issn = {0025-5564},
    shorttitle = {Numerical parameter identifiability and estimability},
    doi = {10.1016/0025-5564(85)90098-7},
    number = {1},
    journal = {Mathematical Biosciences},
    author = {Jacquez, John A. and Greif, Peter},
    month = dec,
    year = {1985},
    pages = {201--227},
}

@article{raue_2009_structural,
    title = {Structural and practical identifiability analysis of partially observed dynamical models by exploiting the profile likelihood},
    volume = {25},
    issn = {1367-4803},
    doi = {10.1093/bioinformatics/btp358},
    number = {15},
    journal = {Bioinformatics},
    author = {Raue, A. and Kreutz, C. and Maiwald, T. and Bachmann, J. and Schilling, M. and Klingmüller, U. and Timmer, J.},
    month = aug,
    year = {2009},
    pages = {1923--1929},
}

@article{miao_2011_identifiability,
    title = {On {Identifiability} of {Nonlinear} {ODE} {Models} and {Applications} in {Viral} {Dynamics}},
    volume = {53},
    issn = {0036-1445},
    doi = {10.1137/090757009},
    number = {1},
    journal = {SIAM Review},
    publisher = {Society for Industrial and Applied Mathematics},
    author = {Miao, Hongyu and Xia, Xiaohua and Perelson, Alan S. and Wu, Hulin},
    month = jan,
    year = {2011},
    pages = {3--39},
}

@article{burth_1999_subset,
    title = {Subset selection for improved parameter estimation in on-line identification of a synchronous generator},
    volume = {14},
    issn = {1558-0679},
    doi = {10.1109/59.744536},
    number = {1},
    journal = {IEEE Transactions on Power Systems},
    author = {Burth, M. and Verghese, G.C. and Velez-Reyes, M.},
    month = feb,
    year = {1999},
    pages = {218--225},
}

@article{cintron-arias_2020_sensitivity,
  title={A sensitivity matrix methodology for inverse problem formulation},
  author={Cintr{\'o}n-Arias, Ariel and Banks, Harvey Thomas and Capaldi, Alex and Lloyd, Alun L},
  journal={Journal of Inverse and Ill-Posed Problems},
  pages={545},
  year={2009}
}

@article{gutenkunst_2007_universally,
    title = {Universally {Sloppy} {Parameter} {Sensitivities} in {Systems} {Biology} {Models}},
    volume = {3},
    issn = {1553-7358},
    doi = {10.1371/journal.pcbi.0030189},
    language = {en},
    number = {10},
    journal = {PLOS Computational Biology},
    publisher = {Public Library of Science},
    author = {Gutenkunst, Ryan N. and Waterfall, Joshua J. and Casey, Fergal P. and Brown, Kevin S. and Myers, Christopher R. and Sethna, James P.},
    month = oct,
    year = {2007},
    pages = {e189},
}

@article{chis_2016_relationship,
    title = {On the relationship between sloppiness and identifiability},
    volume = {282},
    issn = {0025-5564},
    doi = {10.1016/j.mbs.2016.10.009},
    journal = {Mathematical Biosciences},
    author = {Chis, Oana-Teodora and Villaverde, Alejandro F. and Banga, Julio R. and Balsa-Canto, Eva},
    month = dec,
    year = {2016},
    pages = {147--161},
}

@article{aitchison_1958_maximumlikelihood,
    title = {Maximum-{Likelihood} {Estimation} of {Parameters} {Subject} to {Restraints}},
    volume = {29},
    issn = {0003-4851},
    number = {3},
    journal = {The Annals of Mathematical Statistics},
    publisher = {Institute of Mathematical Statistics},
    author = {Aitchison, J. and Silvey, S. D.},
    year = {1958},
    pages = {813--828},
}

@article{gorman_1990_lower,
    title = {Lower bounds for parametric estimation with constraints},
    volume = {36},
    issn = {1557-9654},
    doi = {10.1109/18.59929},
    number = {6},
    journal = {IEEE Transactions on Information Theory},
    author = {Gorman, J.D. and Hero, A.O.},
    month = nov,
    year = {1990},
    pages = {1285--1301},
}

@article{liu_2013_singleimage,
    title = {Single-{Image} {Noise} {Level} {Estimation} for {Blind} {Denoising}},
    volume = {22},
    copyright = {https://ieeexplore.ieee.org/Xplorehelp/downloads/license-information/IEEE.html},
    issn = {1057-7149, 1941-0042},
    doi = {10.1109/TIP.2013.2283400},
    number = {12},
    journal = {IEEE Transactions on Image Processing},
    author = {Liu, Xinhao and Tanaka, Masayuki and Okutomi, Masatoshi},
    month = dec,
    year = {2013},
    pages = {5226--5237},
}

@article{kemajoumbakam_2025_empiricalbayes,
    title = {Empirical {Bayesian} {Image} {Restoration} by {Langevin} {Sampling} with a {Denoising} {Diffusion} {Implicit} {Prior}},
    volume = {67},
    issn = {0924-9907},
    doi = {10.1007/s10851-025-01265-7},
    number = {5},
    journal = {Journal of Mathematical Imaging and Vision},
    publisher = {Springer Science and Business Media LLC},
    author = {Kemajou Mbakam, Charlesquin and Giovannelli, Jean-Francois and Pereyra, Marcelo},
    month = sep,
    year = {2025},
}

@inproceedings{martin_2001_database,
  title = {A database of human segmented natural images and its application to evaluating segmentation algorithms and measuring ecological statistics},
  volume = {2},
  doi = {10.1109/ICCV.2001.937655},
  booktitle = {Proceedings {Eighth} {IEEE} {International} {Conference} on {Computer} {Vision}. {ICCV} 2001},
  author = {Martin, D. and Fowlkes, C. and Tal, D. and Malik, J.},
  month = jul,
  year = {2001},
  pages = {416--423 vol.2},
}

@inproceedings{agustsson_2017_ntire,
  title = {{NTIRE} 2017 {Challenge} on {Single} {Image} {Super}-{Resolution}: {Dataset} and {Study}},
  issn = {2160-7516},
  shorttitle = {{NTIRE} 2017 {Challenge} on {Single} {Image} {Super}-{Resolution}},
  doi = {10.1109/CVPRW.2017.150},
  booktitle = {2017 {IEEE} {Conference} on {Computer} {Vision} and {Pattern} {Recognition} {Workshops} ({CVPRW})},
  author = {Agustsson, Eirikur and Timofte, Radu},
  month = jul,
  year = {2017},
  note = {ISSN: 2160-7516},
  pages = {1122--1131},
}

@article{virtanen2020scipy,
  title={SciPy 1.0: fundamental algorithms for scientific computing in Python},
  author={Virtanen, Pauli and Gommers, Ralf and Oliphant, Travis E and Haberland, Matt and Reddy, Tyler and Cournapeau, David and Burovski, Evgeni and Peterson, Pearu and Weckesser, Warren and Bright, Jonathan and others},
  journal={Nature methods},
  volume={17},
  number={3},
  pages={261--272},
  year={2020},
  publisher={Nature Publishing Group US New York}
}

@book{bouman_2022_foundations,
  title={Foundations of computational imaging: a model-based approach},
  author={Bouman, Charles A},
  year={2022},
  publisher={SIAM}
}
